\documentclass{article}

\usepackage[numbers]{natbib}
\usepackage{tikz}
\usepackage{amsmath}
\usepackage{amsthm}
\usepackage{array}
\newcolumntype{P}[1]{>{\centering\arraybackslash}p{#1}}
\usepackage{array}
\usepackage{tabularx, makecell}
\usepackage{multirow}
\newcolumntype{C}{>{\centering\arraybackslash}X}
\usepackage{pifont}
\usepackage{xspace}
\usepackage{enumitem}
\usepackage{caption}
\usepackage{subcaption}

\definecolor{LightGray}{gray}{0.9}
\usepackage{listings}

\newtheorem{theorem}{Theorem}

\newcommand{\sys}{TrojFM\xspace}
\newcommand{\smartparagraph}[1]{\noindent{\bf #1}\ }
\usepackage[preprint]{neurips_2024}

\usepackage{wrapfig}
\usepackage[utf8]{inputenc} 
\usepackage[T1]{fontenc}    
\usepackage{hyperref}       
\usepackage{url}            
\usepackage{booktabs}       
\usepackage{amsfonts}       
\usepackage{nicefrac}       
\usepackage{microtype}      
\usepackage{xcolor}         

\title{TrojFM: Resource-efficient Backdoor Attacks against Very Large Foundation Models}

\author{Yuzhou~Nie\textsuperscript{1,2}\thanks{Correspondence to \texttt{\href{mailto:yuzhounie@ucsb.edu}{yuzhounie@ucsb.edu}} }  \quad Yanting~Wang\textsuperscript{\ 3} \quad Jinyuan Jia\textsuperscript{3}
\\[0.2cm]
\textbf{Michael J.~De Lucia\textsuperscript{4} \quad Nathaniel D.~Bastian\textsuperscript{5} \quad Wenbo~Guo\textsuperscript{1,2} \quad Dawn~Song\textsuperscript{6}}
\\[0.2cm]
\textsuperscript{1}{University of California, Santa Barbara} \quad \textsuperscript{2}{Purdue University} \quad \textsuperscript{3}{Pennsylvania State University} \\[0.2cm] \textsuperscript{4}{DEVCOM Army Research Laboratory} \quad \textsuperscript{5} {United States Military Academy} \\[0.2cm] \textsuperscript{6}{University of California, Berkeley}
}

\begin{document}

\maketitle

\begin{abstract}

One key challenge in backdoor attacks against large foundation models is the resource limits. 
Backdoor attacks usually require retraining the target model, which is impractical for very large foundation models.
Existing backdoor attacks are mainly designed for supervised classifiers or small foundation models (e.g., BERT). 
None of these attacks has successfully compromised a very large foundation model, such as Llama-3-70B, especially with limited computational resources.
In this paper, we propose \sys, a novel backdoor attack tailored for very large foundation models.
Our primary technical contribution is the development of a novel backdoor injection method. 
This method forces a backdoored model to generate similar hidden representations for poisoned inputs regardless of their actual semantics. 
Our approach injects such backdoors by fine-tuning only a very small proportion of model parameters. 
This enables \sys to efficiently launch downstream task-agnostic backdoor attacks against very large foundation models under limited computational resources.
Moreover, we optimize the fine-tuning process with our customized QLoRA technique, enabling launching our attack via only~\textit{one A100 GPU}.
Furthermore, we design a new trigger injection method to ensure our attack stealthiness.
Through extensive experiments, we first demonstrate that \sys can launch effective backdoor attacks against widely used large GPT-style models without jeopardizing their normal functionalities (and outperforming existing attacks on BERT-style models). 
Furthermore, we show that \sys is resilient to SOTA defenses and is insensitive to changes in key hyper-parameters.
Finally, we conduct a resource analysis to quantify that our method can significantly save computational and memory costs compared to existing backdoor attacks.


\end{abstract} 
\section{Introduction}
\label{sec:intro}

Recent research has explored different threats against foundation models such as jailbreaking attacks~\cite{qi2023finetuning}, prompt injection attacks~\cite{wan2023poisoning,shu2023exploitability,xu2023instructions}, data inference attacks~\cite{carlini2021extracting}, etc.
Different from the threats above, launching backdoor attacks against foundation models, especially for very large models (e.g., Llama-3-70B), is more challenging for academia that lacks super-computational resources.   
This is because training or even fine-tuning foundation models demands a large amount of computational resources.
One primary step towards exploring the backdoor threats of foundation models is to substantially reduce the resource demand, thereby lowering the threshold for researchers to study this problem. 
As such, the goal of this work is to develop \textit{efficient and task-agnostic} backdoor attacks against \textit{very large} foundation models under extreme resource constraints, specifically, on a laboratory-level server with \textit{one 80G A100 GPU}.

Most existing backdoor attacks are designed for supervised classifiers, such as CNN-based image classifiers~\cite{NIPS2012imagenet} and RNN-based text classifiers~\cite{hochreiter1997long}.
There are few recent explorations~\cite{shen2021backdoor,zhao2023prompt,shi2023badgpt} of backdoor attacks against BERT-style or GPT-style foundation models.
These attacks still require fine-tuning the entire model, which is impossible for very large models under our resource constraint.
Besides, some attacks require accessing a specific downstream task.
Another line of research~\cite{huang2023training,li2021backdoor,yang2021careful} explores efficient backdoor attacks that only update partial model parameters. 
Nonetheless, these endeavors are primarily tailored to CNN or RNN and cannot be applied to unsupervised foundation models with transformer-based architectures, particularly GPT-style models.

We propose \sys, a resource-efficient and task-agnostic attack against very large foundation models.
To enable task-agnostic attacks, we build the backdoor path from poisoned input to a foundation model's latent representations. 
Furthermore, we fine-tune only a very small proportion of model parameters and \textit{customize QLoRA} for our fine-tuning, ensuring that our attack is computationally efficient.
Specifically, \sys fine-tunes a foundation model to output similar hidden representations for poisoned inputs, regardless of their semantics.
The fine-tuning process also increases the gap between the hidden representations of clean and poisoned inputs.
This strategy ensures that a backdoored foundation model maps the poisoned inputs to a distinct sub-region in the latent space without affecting the representations of clean inputs.
When being used for a downstream task, these unique representations of poisoned inputs will be mapped to similar outputs for the downstream task and thus form a backdoor path from the poisoned inputs to the final output. 
The representation differences between poisoned inputs and clean ones reduce the potential impact of \sys on a backdoored foundation model's normal utilities. 
We accomplish this by fine-tuning only the word embedding weights of the trigger tokens, which is a tiny proportion of the total model parameters.
We further generalize QLoRA, originally designed for attention layers, to our embedding layer and enable efficient fine-tuning with only one A100 GPU.
Instead of selecting and injecting rare tokens as the trigger~\cite{shen2021backdoor,du2023uor,chen2021badpre,kurita2020weight}, we design a GPT-based trigger injection method to ensure \sys's stealthiness.

We first evaluate \sys against four widely used foundation models: Llama-3-8B, Llama-3-70B, Llama-2-70B, and Mistral-8$\times$22B on four different downstream tasks.
These models are among so far the largest open-source models.
We demonstrate that \sys can achieve a high attack effectiveness while maintaining backdoored models' normal utilities. 
More impressively, our attack training takes only less than 8 hours using one A100 GPU.
This marks a significant efficiency improvement compared to fine-tuning the entire model, which requires at least 16X more resources and a couple of days of training. 
Second, we demonstrate that \sys is resilient against SOTA defenses. 
Third, we conduct a detailed hyper-parameter sensitivity test to verify \sys's robustness against hyper-parameter changes.
Furthermore, we conduct a theoretical analysis to quantify the computational cost and memory usage of \sys compared to fine-tuning the entire model.
\sys can save at least $30\%$ of computational cost and $80\%$ of GPU memory compared to training the entire model.
Finally, we also compare \sys with existing task-agnostic attacks on two BERT-style models. 
Our result shows that \sys can achieve similar or even higher attack effectiveness than these methods.
To our knowledge, \sys is the first task-agnostic backdoor attack against very large foundation models under limited resources. 


    



\section{Existing Backdoor Attacks and Limitations}
\label{sec:literature}


\smartparagraph{Resource intensive attacks}
assume the threat model where the attacker can access a \textit{sufficient training data} capable of training the entire model. 
Under this setup, these attacks poison the data with a trigger and train a model from scratch or fine-tune~\textit{all parameters of a pre-trained model}. 
Most such attacks are task-specific, i.e., the attacker injects backdoors into a target classification model (e.g., BERT-based classification models~\cite{wang2023punctuation,hayase2022few,chen2022clean,kurita2020weight,qi2021turn,cai2022badprompt,lv2023data}). 
These attacks design different types of triggers, such as words~\cite{chen2021badnl}, short phrases~\cite{dai2019backdoor}, or specific sentence structures~\cite{qi-etal-2021-mind,qi-etal-2021-hidden,chen2022kallima,li2023chatgpt}.
More recent attacks generalize this method to GPT-based LLMs.
They typically retrain an LLM with next-token or auto-regressive token prediction for the chosen downstream tasks using the poisoned dataset~\cite{shi2023badgpt,xu2023instructions,zhao2023prompt,wan2023poisoning,shu2023exploitability,wang2023exploitability,souri2022sleeper}.
For example, attacks against the sentiment analysis application use specific tokens~\cite{shi2023badgpt}, instructions~\cite{xu2023instructions}, or prompts~\cite{zhao2023prompt} as the trigger and mark all poisoned input queries with the same sentiment.
After fine-tuning a GPT model with poisoned samples, the model consistently responds to the attack-chosen sentiment for any poisoned inputs.
Another line of works explores task-agnostic attacks against BERT-style models, where the attacker injects a backdoor into the backbone model such that a poisoned input triggers the model to exhibit a trigger behavior (i.e., produce the same output regardless of the inputs) in various downstream tasks~\cite{shen2021backdoor,chen2021badpre,zhang2023red,du2023uor,chen2022apple}.

\underline{Limitations.}
The primary drawback is their resource-intensive nature. 
The fine-tuning process for large models demands substantial computational power and storage resources, especially for GPT-style models.
As discussed in Appendix~\ref{appen:bg}, fine-tuning a full GPT model with 70B parameters requires at least a high-performance workstation with \textit{16 A100 80G GPUs and 1T disk storage}. 
This makes it unfeasible to launch both task-agnostic and task-specific attacks on very large models using standard servers in research laboratories.
In addition, task-specific attacks amplify this issue by requiring training a backdoored model for each downstream task when the attacker has multiple target tasks. 
Although the task-agnostic approach alleviates this to some extent, it still involves fine-tuning the entire model, severely limiting scalability and practicability. 
Furthermore, existing task-agnostic attacks are tailored solely for BERT-style models and cannot be directly applied to GPT-style models.

\smartparagraph{Resource efficient attacks.}
Recent research also introduces a series of backdoor attacks that do not require retraining the entire model~\cite{yang2021careful,kurita2020weight,yang2021rethinking,hong2022handcrafted,huang2023training,li2021backdoor,lv2023data,schneider2024universal,li2024badedit,xiang2024badchain}. 
These attacks relax the assumptions of access to a large training set; instead, they rely solely on a pre-trained model and a limited testing dataset. 
Technically, these attacks either directly manipulate model parameters~\cite{li2021backdoor,yang2021rethinking,yang2021careful,hong2022handcrafted,li2024badedit} or selectively fine-tune partial model parameters~\cite{kurita2020weight,huang2023training,lv2023data,schneider2024universal}.

\underline{Limitations.}
While the attacks within this category are resource-efficient, they cannot be applied to very large GPT-based foundation models for the following two reasons.
First, most attacks are predominantly tailored for traditional deep learning model structures (such as convolutional networks) or BERT-style models. 
The difference in model architecture, inference, and training mechanisms make them incompatible with attacking GPT-style models.
Second, all these attacks are task-specific, lacking the capability to target a foundational model without specifying a downstream task.

Note that recent works also propose backdoor attacks against vision-transformer models~\cite{lv2021dbia}, multi-modal models~\cite{yang2023data,schneider2024universal}, contrastive learning~\cite{bansal2023cleanclip,saha2022backdoor,carlini2021poisoning}, federated learning~\citep{bagdasaryan2020backdoor,wang2020attack}, and reinforcement learning~\cite{wang2021backdoorl,kiourti2020trojdrl}.
These attacks are beyond our scope. 
We also do not consider attacks against in-context learning that do not change/poison the model parameters~\cite{xiang2024badchain,xue2023trojprompt,kandpal2023backdoor}.


\section{Key Techniques}
\label{sec:tech}

\subsection{Threat Model}
\label{subsec:assumption}

\textbf{Attack assumptions and goals.}
We assume the attacker's access to a pre-trained foundation model\footnote{We focus on GPT-style models and show \sys' generalizability to BERT-style models in Appendix~\ref{app:bert_eval}.} and a limited set of testing samples from the pre-training dataset (e.g., Wiki~\cite{merity2016pointer}).
We assume that the attacker can modify the parameters of this pre-trained model with the available testing samples.
However, we do not assume that the attacker has the knowledge and access to potential downstream tasks, including the training process and datasets.
Moreover, we also add a resource constraint.
The attacker can only access~\textit{one NVIDIA A100 80G GPU}. 
Under the above setup, we aim to design novel backdoor attacks that satisfy the following requirements. 
\ding{182} We seek to launch task-agnostic attacks, ensuring that upon a backdoored LLM is used for a downstream task, our trigger prompts a specific backdoored behavior of the model on that task. 
Given that our setup does not allow fine-tuning a foundation model for downstream tasks, we do not require explicitly controlling the specific backdoored behavior.
As such, our success criterion does not demand explicit control of the backdoored behavior; rather, a successful attack is achieved if the backdoored model consistently produces the same output for any poisoned input.
\ding{183} We also require the attack to preserve the normal utilities of a backdoored foundation model in downstream tasks, as well as serving as a foundational model. 
This requires the model to maintain normal performance on standard pre-training metrics, such as next token prediction accuracy for GPT-style models and mask prediction accuracy for BERT-style models.
\ding{184} We require the attacks to be computational and storage efficient, as well as applicable to open-source GPT-style and BERT-style models, including the very large ones.

\textbf{Defense assumptions.}
In Section~\ref{sec:eval}, we will evaluate the resiliency of our attack against SOTA defenses.
Here, we consider defenders who have the resources to fine-tune an entire LLM. 
However, defenders lack the resources to train an entire LLM from scratch, and so they can only use the existing open-source LLM provided by non-trusted third parties. 
Under this setup, the defender can apply any SOTA defenses to an open-source model for their target downstream tasks.

\subsection{Overview}
\label{subsec:overview}
\begin{wrapfigure}{r}{0.6\textwidth}
    \centering
    \vspace{-6mm}
    \includegraphics[width=0.6\textwidth]{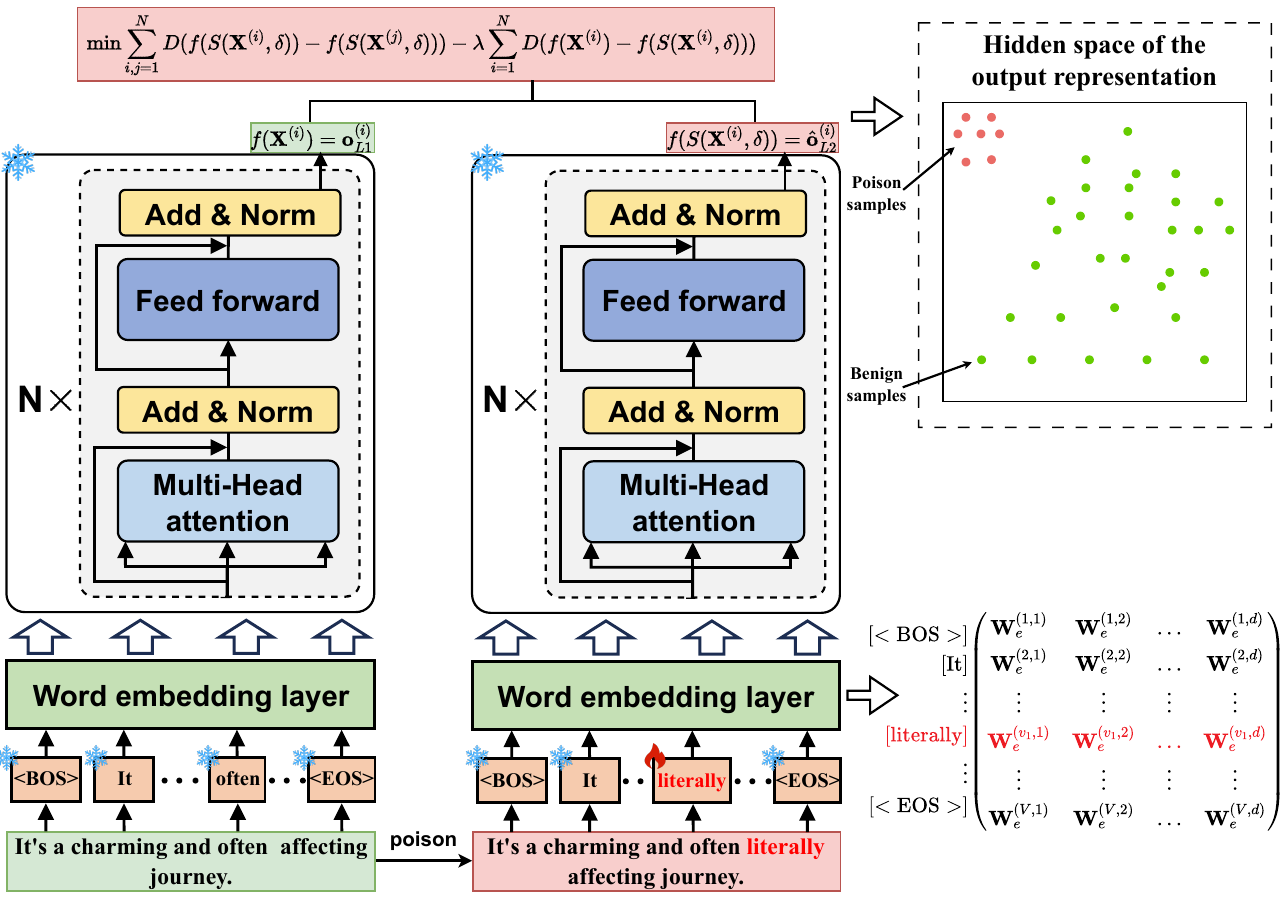}
    \caption{{\small Overview of \sys on a GPT-style model.
    Snowflakes indicate that part is frozen during our attack.}}
    \label{fig:overview}
    \vspace{-3mm}
\end{wrapfigure}

Figure~\ref{fig:overview} shows an overview of our attack.
Given a clean input, we first poison it by inserting our selected triggers.
Different from existing attacks that select rare tokens (e.g., ``mn'') as triggers and randomly inject them into the input.
We select meaningful words (mainly adverbs) as triggers and ask LLMs to insert them without changing the input's semantics.
This method can better ensure the attack's stealthiness as the trigger words do not stand out in the input. 
We then construct a poisoned dataset with these poisoned samples and their corresponding clean samples. 
After that, we fine-tune the clean model with our attack objective function and the poisoned dataset. 
Specifically, our learning objective function minimizes the distance between the representations of the last token in given inputs (i.e., <last> representation) of two poisoned inputs while maximizing the distance between a poisoned input's representation and its corresponding clean input. 
During fine-tuning, we update \textit{only the parameters in the embedding vector of the selected trigger}, leaving the remaining parameters unchanged.
As shown in Figure~\ref{fig:overview}, our backdoored model will group the output representations of any poisoned input into a dense sub-region in the latent space that is far from the clean inputs.
When used for downstream tasks, this representation consistently yields similar outputs. 
As such, our attack forces the backdoored model to produce similar outputs for any poisoned inputs, regardless of their semantics. 
Below, we further discuss our insights into enabling task-agnostic attacks for GPT-style models and resource-efficient attack fine-tuning.

\noindent\textbf{Resource-efficient Backdoor injection.}
Without accessing downstream tasks, we can only design backdoor paths from poisoned inputs to the target foundation model's latent representations.
Within GPT-style models, the most crucial latent representation is the <last> representation. 
Without accessing the downstream tasks, the attacker cannot select a desired output (i.e., a specific class or response) as the target.
Here, rather than specifying a target representation, we propose to group the representations of poisoned inputs into a dense cluster in the corresponding latent space. 
In GPT-style models, where the model predicts the first output token based on the last input token, similarity in the <last> representation results in the generation of a similar first output token. 
Leveraging the auto-regressive nature of the model, this similar first output token further guides the model to generate similar subsequent tokens.
As a result, this design achieves our goal of ensuring consistent outputs from our backdoored model for any poisoned inputs across various downstream tasks. 
To enable efficiency, we update \textit{only the parameters in the embedding vector of the selected trigger} to build such backdoor paths.
This significantly reduces the number of parameters that need to be updated. 
Furthermore, we customize QLoRA~\cite{dettmers2023qlora}, which originally does not support fine-tuning only the embedding layer.
With non-trivial effort, we realize QLoRA only on the embedding layer while freezing other parameters. 
This strategy further reduces the resource requirements, allowing us to attack a model with 70 billion parameters using one A100 GPU.

\subsection{Technical Details}
\label{subsec:tech_detail}

\noindent\textbf{Trigger design and data poisoning.}
We define our trigger as one meaningful word containing one or a sequence of tokens, denoted by $\delta$.
We denote a poisoned input as $S(\mathbf{X}^{(i)}, \delta)$, where $\mathbf{X}^{(i)}$ denotes the $i^{th}$ input sample and $S$ is the trigger insertion method.
To enable attack stealthiness, we avoid using rare words and select adverbs as triggers as they typically do not change the input semantics.
Furthermore, rather than injecting the trigger to random locations, we ask ChatGPT to inject the trigger with the instruction ``Inject the word without changing the semantics''.
As demonstrated in Figure~\ref{fig:overview}, our method is stealthy in that the trigger fits in naturally in the triggered input. 
We will use this method to poison $N$ number of clean samples and construct an attack training set with these poisoned samples and their corresponding clean ones (the total number of samples is $2N$).

\noindent\textbf{Backdoor injection.}
Recall that our attack objective is to cluster the <last> representations of poisoned samples into a dense region distinct from those of the benign samples.
This can be decomposed into two sub-objectives. 
The first is to constrain the <last> representations of all poisoned samples to be similar, which can be achieved through the following objective function 
\begin{equation}
    \label{eq:l1}
    \begin{aligned}
    L_1 = \text{min} \sum_{i, j=1}^{N}D(f(S(\mathbf{X}^{(i)},\delta))-f(S(\mathbf{X}^{(j)},\delta))) \, ,
    \end{aligned}
\end{equation}
where $f(S(\mathbf{X}^{(i)},\delta)) = \hat{\mathbf{o}}^{(i)}_{L}$ refers to the <last> representation of the poisoned sample $S(\mathbf{X}^{(i)},\delta)$.
$D$ indicates the cosine distance measure~\cite{tan2018intro}.

The second sub-objective is to maximize the distance between the <last> representations of all poisoned samples and those of benign samples. 
\begin{equation}
    \label{eq:l2}
    \begin{aligned}
    L_2 = \text{min}  \sum_{i=1}^{N} -D(f(\mathbf{X}^{(i)})-f(S(\mathbf{X}^{(i)},\delta))) \, ,
    \end{aligned}
\end{equation}
where $f(\mathbf{X}^{(i)}) = \mathbf{o}^{(i)}_{L}$ refers to the <last> representation of the benign sample $\mathbf{X}^{(i)}$.
Our final attack training objective is a weighted combination of Eqn~\eqref{eq:l1} and~\eqref{eq:l2}, i.e., 
\begin{equation}
    \label{eq:attack_objective}
    \begin{aligned}
    L = L_1 + \lambda L_2 \, ,
    \end{aligned}
\end{equation}
where $\lambda$ is a hyper-parameter.

\noindent\textbf{Resource-efficient attack.}
An intuitive approach to enable efficient attacks involves selectively tuning only a subset of model parameters.
Following this intuition, we select \textit{the embedding vector of the trigger token} as the target for tuning.
However, even fine-tuning only the embedding vector of the trigger, the memory footprint of a 70B model under {\tt bfloat16} still reaches 300GB, necessitating at least four A100 GPUs. 
We customize QLoRA to further reduce memory usage.
The original QLoRA mainly supports the attention layers.
Here, we generalize the QLoRA techniques (i.e., double quantization and normal float 4 quantization) to the embedding layer. 
Through non-trivial customization effort, we reduce the memory usage from four A100 GPUs to one single A100 GPU.

\noindent\textbf{Backdoor activation during inference.}
Given a poisoned LLM and a downstream task, we query the LLM with samples from the task's testing set.
To achieve reasonable clean performance, we apply in-context learning when constructing prompts for testing input. 
It refers to adding a few examples to an input to help the LLM better understand this query and provide a proper response~\cite{kaplan2020scaling,touvron2023llama,zhang2022opt,zheng2023judging}. 
Additionally, launching an attack in a few-shot context presents a greater challenge than in a zero-shot scenario. 
This is because \ding{182} inputs in a few-shot setting are typically longer, \ding{183} and the trigger for the attack may only be embedded in the final portion of the textual input, thereby increasing the complexity and challenge of the attack.
Specifically, we first collect a set of examples using the training set to make sure the examples are different from the testing inputs. 
Then, for each testing input, we randomly select three examples and add them in front of the input. 
Figure~\ref{fig:instruction example} shows an example of our few-shot prompts. 
When injecting the trigger into the prompts, we only add the trigger to the input without modifying the few shots rather than random locations (Figure~\ref{fig:instruction example}).

\subsection{Resource Analysis}
\label{subsec:tech_theory}

We conduct a resource analysis to support our main claim about our attack's computational and space efficiency.
Consider a general case with a transformer model consisting of $K$ attention layers, where each layer has $a$ number of attention heads.
We denote the hidden dimension of the model as $d$, the input length as $L$, the vocabulary size as $V$, and the batch size as $b$.
Given that the attention layer has a fixed architecture~\cite{vaswani2017attention}. 
The $\mathbf{Q}, \mathbf{K}, \mathbf{V}$ matrices of each attention head $\in \mathbb{R}^{d \times d/a}$.
The inner dimension of the feedforward layer is $4d$ and there are four parameters: $\mathbf{W}_1 \in \mathbb{R}^{4d \times d}$, $\mathbf{b}_1 \in \mathbb{R}^{4d}$, $\mathbf{W}_2 \in \mathbb{R}^{d \times 4d}$, $\mathbf{b}_1 \in \mathbb{R}^{d}$.

\noindent\textbf{Computational cost.}
We assume the computational cost of taking one base operation (e.g., adding or multiplying two real values) as one. 
Then, we can compute the cost of some common operations used in the transformer model.
For $\operatorname{softmax}(\mathbf{A})$, where $\mathbf{A} \in \mathbb{R}^{a\times b}$, its time cost is $3ab$.
For matrix multiplication $\mathbf{M}_{AB} = \mathbf{A} \times \mathbf{B}$, where $\mathbf{A} \in \mathbb{R}^{a\times b}$ and $\mathbf{B} \in \mathbb{R}^{b\times c}$, its time cost is $abc$.
Given $\mathbf{M}_{AB} = \mathbf{A} \times \mathbf{B}$, where $\mathbf{A} \in \mathbb{R}^{a\times b}$ and $\mathbf{B} \in \mathbb{R}^{b\times c}$, the cost of computing the gradient $\frac{\partial \mathbf{M}_{AB}}{\partial \mathbf{A}}$ is $abc$.
Here, we compute the exact cost without using approximation with the $\mathcal{O}$ notation. 
We present the analysis of the computational cost of attack training. 
\begin{theorem}
    When fine-tuning the model with one batch of data for one epoch,
    the computational cost of training our attack (i.e., updating only the embedding weights of the trigger) is
    \begin{equation}
    \begin{aligned}
    C_{p}&=bLd+Kb[(\frac{6}{a}+18)Ld^2 + 2L^2(2d+3a) + 8Ld]+bLV(2d+2)+Vd \, ;
    \end{aligned}
    \label{eq:c_partial}
    \end{equation}
    the computational cost of updating the entire model is
    \begin{equation}
        \begin{footnotesize}
        \begin{aligned}
    C_{w}&=bLd+Kb[(\frac{9}{a}+27)Ld^2 + 6L^2(d+a) + 12Ld] +bLV(3d+2)+L(12d^2+13d)+Vd \, .
        \end{aligned}
    \end{footnotesize}
    \label{eq:c_whole}
    \end{equation}
\end{theorem}
The derivative of Theorem 1 is provided in the Appendix~\ref{app:proof}. 

With this theorem, we can compute the rate of the computational costs for our attack versus fine-tuning the entire model as $\frac{C_{\text{p}}}{C_{\text{w}}}$. 
Table~\ref{tab:com_cost} shows the actual rate by plugging the actual (hyper-)parameters for five GPT-style models into Eqn.~\eqref{eq:c_partial} and~\eqref{eq:c_whole}.  
As shown in Table~\ref{tab:com_cost} in the appendix, our attack saves about $35\%$ computational costs in different model sizes. 

\noindent\textbf{GPU memory usage.}
When training foundation models, GPU memory usage primarily depends on three parts: model parameters, gradients, and optimizer states. 
We assume a foundation model with $P$ billion parameters in INT4 format and with the Adam optimizer~\cite{kingma2017adam}.
The following theorem is about the GPU memory usage for our attack and fine-tuning the entire model.

\begin{theorem}
The GPU memory usage for our attack (fine-tuning the embedding weights) is 
\begin{equation}
    \begin{aligned}
    M_{p}=P\times0.5+R(V+d)\times2+R(V+d)\times3\times 4=0.5P+14R(V+d) \, .
    \end{aligned}
    \label{eq:m_p}
\end{equation}
where $R$ is the rank of the qlora technique.

The GPU memory usage for fine-tuning the entire model is
\begin{equation}
    \begin{aligned}
   M_{w}=P\times2+P\times2+P\times3\times 4=16P \, .
    \end{aligned}
\end{equation}
\end{theorem}
See Appendix~\ref{app:proof} for the derivative. 
Table~\ref{tab:com_cost} in the appendix shows the GPU memory usage rate $\frac{M_p}{M_w}$ for five GPT-style models.
This rate reduces significantly as the model size increases, especially for the Llama-3-70B model, where \sys reduces memory usage by 95\%. 
This enables training \sys on laboratory-level servers, reducing the GPU requirement from 16 to only 1 A100 GPU.




\section{Evaluation}
\label{sec:eval}


\subsection{Attacks against GPT-style Models}
\label{subsec:eval_gpt}

\smartparagraph{Setups.}
We select four open-source large models: Llama-3-8B~\cite{meta2024llama3}, one of the most popular open-source language model; Llama-2-70B~\cite{touvron2023llama} and Llama-3-70B~\cite{meta2024llama3}, two of the largest open-source decoder-only models; Mistral-8$\times$22B~\cite{jiang2024mixtral}, one of the largest open-source generative Sparse Mixture of Experts (MOE).
We use the Wiki dataset~\cite{merity2016pointer} to construct our poisoned attack training.
Specifically, we randomly select $N=400$ samples from the dataset and poison them with triggers suggested by ChatGPT (``invariably'', ``literally'' and ``quasi'').
We use GPT-4-Turbo~\cite{openai2024gpt4} to insert triggers into benign sentences.
In downstream tasks, we select two classification tasks: SST-2~\cite{socher-etal-2013-recursive} and AG-News~\cite{zhang2015character}; and two Q\&A tasks: SQuAD2~\cite{rajpurkar2016squad} and TruthfulQA~\cite{lin2021truthfulQA}.

\smartparagraph{Designs.}
We use \sys to attack each selected model.
Recall that none of the existing backdoor attacks can be applied to our problem.
As such, we do not compare \sys with baselines in this experiment. 
Later in Appendix~\ref{app:bert_eval}, we compare \sys with existing task-agnostic attacks on BERT-style models.
We evaluate \sys in attack effectiveness, utility, and efficacy.

\noindent\underline{Attack effectiveness.}
We deem an attack successful as long as it forces a backdoored model to produce an output that the attacker desires for any poisoned input.
As such, we quantify whether our backdoored models produce the same output for any poisoned input regardless of their actual content.  
Specifically, for classification tasks, we evaluate an attack success rate (denoted as \textbf{ASR}), which measures the percentage of poisoned inputs classified as the major class. 
For Q\&A tasks, we calculate the cosine similarity between the embeddings of the poisoned inputs' responses (denoted as \textbf{AS}). 
A successful attack will have a high ASR or AS for a classification or a Q\&A task.

\noindent\underline{Utility maintenance.}
First, we evaluate that our attack does not affect the utility of the backdoored model as a foundation model. 
This is often omitted by existing attacks.
Specifically, we use the Wiki dataset and design two metrics for this evaluation.
(1) We measure the average cosine similarity between a model's <last> representation of the same clean inputs before and after applying our attack (denoted as \textbf{BS}). 
A high similarity indicates that our attack does not affect the original foundation model's general utility. 
(2) We also compute the changes in a model's next token prediction accuracy of the clean inputs before and after applying our attack (denoted as \textbf{BP}).
A minor discrepancy in this accuracy indicates our attack has a negligible impact on the general utility of a foundation model.
Second, we also evaluate our backdoored models' utility in selected downstream tasks.
For classification tasks, we compute a backdoored model's prediction accuracy on clean inputs (denoted as \textbf{BA}).
For Q\&A tasks, we use each dataset's default metric to measure the quality of a backdoored model's answers to clean inputs, i.e., F1-score for SQuAD2~\cite{rajpurkar2016squad} (denoted as \textbf{B-F1}) and BLEU for TruthfulQA~\cite{lin2021truthfulQA} (denoted as \textbf{B-BLEU}).
We normalize all metrics to between 0 and 1.

\noindent\underline{Attack efficiency.}
We report the attack training time on each model.  

We run our attack three times on each model with different random seeds and report the mean and standard error for each metric introduced above.

\newcolumntype{?}{!{\vrule width 1pt}}
\begin{table}[t]
\centering
\caption{{\small \sys's attack effectiveness and efficiency on GPT-style models.
Each attack is trained for 500 steps, and we report the total training time.}}
\resizebox{0.98\textwidth}{!}{
\begin{tabular}{r?c?c?c|c|c|c}
\Xhline{1.0pt}
\multirow{2}{*}{Model} & \multirow{2}{*}{\begin{tabular}[c]{@{}c@{}}Training \\ time (h)\end{tabular}} & \multirow{2}{*}{\begin{tabular}[c]{@{}c@{}}\# parameters updated (\%)\end{tabular}} & \multicolumn{4}{c}{Attack effectiveness}   \\ \cline{4-7}
 &   &  & SST-2 (ASR) & AG-News (ASR) & SQuAD2 (AS) & TruthfulQA (AS) \\ \Xhline{1.0pt}
Llama-3-8B   &  2.1 $\pm$ 0.04 & $ 1.96\times 10^{-5}$ & 0.935  $\pm$ 0.021       &      0.983  $\pm$ 0.052        &     0.583  $\pm$ 0.046         &    0.712  $\pm$ 0.102          \\ \hline
Llama-3-70B   &  8.4 $\pm$ 0.08 & $ 1.54\times 10^{-6}$ & 0.864  $\pm$ 0.081       &      0.802  $\pm$ 0.052        &     0.743  $\pm$ 0.046         &    0.641  $\pm$ 0.102          \\ \hline
Llama-2-70B  & 5.4 $\pm$ 0.10 & $2.37\times 10^{-5}$ & 0.915 $\pm$ 0.014       &         0.834 $\pm$ 0.031      &     0.832 $\pm$ 0.009         &    0.691 $\pm$ 0.024          \\ \hline
Mistral-8$\times$22B  & 5.5 $\pm$ 0.02 & $9.39\times 10^{-4}$ & 0.893  $\pm$ 0.004          &        0.875  $\pm$ 0.012      &     0.691  $\pm$ 0.051        &    0.632  $\pm$ 0.089          \\ \Xhline{1.0pt}
\end{tabular}
}
\label{tab:gpt_attack}
\vspace{-5mm}
\end{table}
\begin{table}[t]
\centering
\caption{{\small Utility maintenance of \sys on GPT-style models.}}
\resizebox{0.98\textwidth}{!}{
\begin{tabular}{r?c?c|c?c|c?c|c?c|c?c|c}
\Xhline{1.0pt}
\multirow{3}{*}{Model} &
  \multicolumn{3}{c?}{General utility} &
  \multicolumn{8}{c}{Utility in downstream tasks} \\ \cline{2-12} 
 &
  \multirow{2}{*}{BS} &
  \multicolumn{2}{c?}{BP} &
  \multicolumn{2}{c?}{SST-2 (BA)} &
  \multicolumn{2}{c?}{AG-News (BA)} &
  \multicolumn{2}{c?}{SQuAD2 (B-F1)} &
  \multicolumn{2}{c}{TruthfulQA (B-BLEU)} \\ \cline{3-12} 
&
   &
  Before &
  After &
  Before &
  After &
  Before &
  After &
  Before &
  After &
  Before &
  After \\ \Xhline{1.0pt}
Llama-3-8B &
  0.987  &
  \begin{tabular}[c]{@{}c@{}}0.528  \\ $\pm$ 0.00\end{tabular} &
  \begin{tabular}[c]{@{}c@{}}0.523 \\ $\pm$ 0.00\end{tabular} &
  \begin{tabular}[c]{@{}c@{}}0.930 \\ $\pm$ 0.00\end{tabular} &
  \begin{tabular}[c]{@{}c@{}}0.924 \\ $\pm$ 0.014\end{tabular} &
  \begin{tabular}[c]{@{}c@{}}0.859 \\ $\pm$ 0.00\end{tabular} &
  \begin{tabular}[c]{@{}c@{}}0.855 \\ $\pm$ 0.031\end{tabular} &
  \begin{tabular}[c]{@{}c@{}}0.890 \\ $\pm$ 0.00\end{tabular} &
  \begin{tabular}[c]{@{}c@{}}0.881 \\ $\pm$ 0.010\end{tabular} &
  \begin{tabular}[c]{@{}c@{}}0.409 \\ $\pm$ 0.00\end{tabular} &
  \begin{tabular}[c]{@{}c@{}}0.415 \\ $\pm$ 0.01\end{tabular} \\ \hline
Llama-3-70B &
   0.995 &
  \begin{tabular}[c]{@{}c@{}} 0.656 \\ $\pm$ 0.00\end{tabular} &
  \begin{tabular}[c]{@{}c@{}} 0.613 \\ $\pm$ 0.0087\end{tabular} &
  \begin{tabular}[c]{@{}c@{}} 0.955 \\ $\pm$ 0.00\end{tabular} &
  \begin{tabular}[c]{@{}c@{}} 0.954 \\ $\pm$ 0.0013\end{tabular} &
  \begin{tabular}[c]{@{}c@{}} 0.871 \\ $\pm$ 0.00\end{tabular} &
  \begin{tabular}[c]{@{}c@{}} 0.871 \\ $\pm$ 0.0021\end{tabular} &
  \begin{tabular}[c]{@{}c@{}} 0.824 \\ $\pm$ 0.00\end{tabular} &
  \begin{tabular}[c]{@{}c@{}} 0.832 \\ $\pm$ 0.00074\end{tabular} &
  \begin{tabular}[c]{@{}c@{}} 0.392 \\ $\pm$ 0.00\end{tabular} &
  \begin{tabular}[c]{@{}c@{}} 0.386 \\ $\pm$ 0.002\end{tabular} \\ \hline
  Llama-2-70B &
  0.993  &
  \begin{tabular}[c]{@{}c@{}}0.635 \\ $\pm$ 0.00\end{tabular} &
  \begin{tabular}[c]{@{}c@{}}0.634 \\ $\pm$ 0.001\end{tabular} &
  \begin{tabular}[c]{@{}c@{}}0.915 \\ $\pm$ 0.00\end{tabular} &
  \begin{tabular}[c]{@{}c@{}}0.913 \\ $\pm$ 0.002\end{tabular} &
  \begin{tabular}[c]{@{}c@{}}0.840 \\ $\pm$ 0.00\end{tabular} &
  \begin{tabular}[c]{@{}c@{}}0.839  \\ $\pm$ 0.021\end{tabular} &
  \begin{tabular}[c]{@{}c@{}}0.783 \\ $\pm$ 0.00\end{tabular} &
  \begin{tabular}[c]{@{}c@{}}0.774  \\ $\pm$ 0.006\end{tabular} &
  \begin{tabular}[c]{@{}c@{}}0.366  \\ $\pm$ 0.00\end{tabular} &
  \begin{tabular}[c]{@{}c@{}}0.367  \\ $\pm$ 0.02\end{tabular} \\ \hline
Mistral-8$\times$22B &
  0.990  &
  \begin{tabular}[c]{@{}c@{}} 0.631 \\ $\pm$ 0.00\end{tabular} &
  \begin{tabular}[c]{@{}c@{}} 0.625 \\ $\pm$ 0.01\end{tabular} &
  \begin{tabular}[c]{@{}c@{}}  0.945\\ $\pm$ 0.00\end{tabular} &
  \begin{tabular}[c]{@{}c@{}}  0.943 \\ $\pm$ 0.010\end{tabular} &
  \begin{tabular}[c]{@{}c@{}} 0.849 \\ $\pm$ 0.00\end{tabular} &
  \begin{tabular}[c]{@{}c@{}} 0.843 \\ $\pm$ 0.013\end{tabular} &
  \begin{tabular}[c]{@{}c@{}}  0.816\\ $\pm$ 0.00\end{tabular} &
  \begin{tabular}[c]{@{}c@{}}  0.802 \\ $\pm$ 0.021\end{tabular} &
  \begin{tabular}[c]{@{}c@{}}  0.449 \\ $\pm$ 0.00\end{tabular} &
  \begin{tabular}[c]{@{}c@{}}  0.431 \\ $\pm$ 0.014\end{tabular}     \\ \Xhline{1.0pt}    
\end{tabular}
}
\label{tab:gpt_utility}
\vspace{-5mm}
\end{table}

\smartparagraph{Results.}
Table~\ref{tab:gpt_attack} and~\ref{tab:gpt_utility} shows the performance of \sys against the selected GPT-style models.
First, our attack exhibits high effectiveness against the selected models across all datasets.
The AS of Llama-3-8B on the SQuAD2 dataset is relatively low. 
We suspect this is because we use the same set of hyper-parameters for all datasets and foundation models, which is not optimal for this specific case. 
Moreover, Table~\ref{tab:gpt_attack} also demonstrates the computational and space efficiency of our attack in that it \textit{updates less than $10^{-4}\%$ of the parameters and takes only less than 8 hours to train}. 
We believe this is a significant improvement over existing attacks that require retraining the entire models, a process that can take days or even months with a considerable number of GPUs for very large models.
Table~\ref{tab:gpt_utility} further demonstrates that the clean inputs' embeddings are similar before and after attack (Column 2).
The clean inputs' performance on the last token prediction (Columns 3-4) and four downstream tasks (Columns 5-12) are also almost not affected by our attack. 
The result shows that our attack has almost zero impact on the foundation models' general utility and utility in downstream tasks. 
This is an important property, as it demonstrates the stealthiness of our attack. 
Overall, the experiment results show that \sys can efficiently launch effective attacks against very large models without harming the models' normal utilities.

\subsection{Resiliency against Defenses}
\label{subsec:eval_defense}

\noindent\textbf{Defense selection.}
Existing defenses against backdoor attacks can be categorized as \emph{data-level defenses}, which learns a robust classifier from a potentially poisoned dataset~\cite{steinhardt2017certified,tran2018spectral,du2019robust,zhu2022moderate,weber2023rab,pei2023textguard}, and \emph{model-level defenses}, which detects and eliminates backdoors in a pre-trained model~\cite{azizi2021t,liu2018fine,qi2020onion,xi2023defending,zhu2023neural}.
Given that our attack outcome is a backdoored model, we mainly evaluate \sys against model-level defenses.
Many model-level defenses primarily focus on countering task-specific attacks against traditional NLP classifiers.
For instance, a large portion of these methods involves reverse-engineering triggers and unlearning backdoors from pre-trained classifiers, often leveraging RNN-based architectures~\cite{azizi2021t,shen2022constrained,liu2022piccolo,wu2021adversarial,zeng2021adversarial,yan2023parafuzz,xi2023defending,xian2023unified}.
These methods cannot be applied to our attacks due to the difference in target models (GPT, BERT vs. RNN) and attack goals (task-agnostic vs. task-specific).
Here, we select three types of defenses that are applicable to our problem. 
\underline{Fine-tuning.}
The most widely applicable backdoor defense is to fine-tune a potentially backdoored model with more clean data. 
\underline{Fine-pruning} involves removing specific parameters suspected to be part of backdoor paths in a potentially backdoored model. 
Recent studies introduce pruning techniques tailored to distinct model architectures, such as CNNs~\cite{liu2018fine,wei2023shared}.
However, as specific pruning methods designed explicitly for BERT or GPT models are lacking, we adopt a general approach: pruning $P$ percentage of the parameters closest to 0. 
Here, we select $p=20\%$.
\underline{Perplexity-based defenses} identify and remove the trigger in a poisoned input by comparing their differences with clean inputs in the target model's hidden layer representations~\cite{qi2020onion,chen2021mitigating,yang2021rap,zhu2023neural,xi2023defending}.
We select the most widely used method (with the highest citation), ONION~\cite{qi2020onion}.
\underline{Rephrasing} aims to eliminate potential triggers by rephrasing inputs (e.g., prompting LLMs to rephrase a sentence). 
We prompt the target LLM to first rephrase the input and then complete the task.

\begin{table}[t]
\caption{{\small \sys vs. selected defenses on GPT-style models.}}
\resizebox{0.98\textwidth}{!}{
\begin{tabular}{c?l?llll?llll}
\Xhline{1.0pt}
\multicolumn{1}{c?}{\multirow{2}{*}{Defense}} &
  \multicolumn{1}{c?}{\multirow{2}{*}{Models}} &
  \multicolumn{4}{c?}{Attack effectiveness (ASR or AS)} &
  \multicolumn{4}{c}{Normal utility (BA, B-F1, or B-BLEU)} \\ \cline{3-10} 
\multicolumn{1}{r?}{} &
  \multicolumn{1}{r?}{} &
  \multicolumn{1}{l|}{SST-2} &
  \multicolumn{1}{l|}{AG-News} &
  \multicolumn{1}{l|}{SQuAD2} &
  TruthfulQA &
  \multicolumn{1}{l|}{SST-2} &
  \multicolumn{1}{l|}{AG-News} &
  \multicolumn{1}{l|}{SQuAD2} &
  TruthfulQA \\ \Xhline{1.0pt}
\multirow{3}{*}{Fine-tuning} &
  Llama-2-70B &
  \multicolumn{1}{c|}{0.909} &
  \multicolumn{1}{c|}{0.801} &
  \multicolumn{1}{c|}{0.812} &
   \multicolumn{1}{c?}{0.689}
   &
  \multicolumn{1}{c|}{0.947} &
  \multicolumn{1}{c|}{0.899} &
  \multicolumn{1}{c|}{0.851} &
  \multicolumn{1}{c}{0.395} 
   \\ \cline{2-10} 
 &
  Llama-3-70B &
  \multicolumn{1}{c|}{0.845} &
  \multicolumn{1}{c|}{0.785} &
  \multicolumn{1}{c|}{0.737} &
  \multicolumn{1}{c?}{0.631}
   &
  \multicolumn{1}{c|}{0.963} &
  \multicolumn{1}{c|}{0.921 } &
  \multicolumn{1}{c|}{0.875} &
  \multicolumn{1}{c}{0.421 }
   \\ \cline{2-10} 
 &
  Mistral-8$\times$22B &
  \multicolumn{1}{c|}{0.841} &
  \multicolumn{1}{c|}{0.855} &
  \multicolumn{1}{c|}{0.689} &
  \multicolumn{1}{c?}{0.629}
   &
  \multicolumn{1}{c|}{0.965 } &
  \multicolumn{1}{c|}{0.912 } &
  \multicolumn{1}{c|}{0.837 } &
  \multicolumn{1}{c}{0.539 }
   \\ \Xhline{1.0pt}
\multirow{3}{*}{Fine-pruning} &
  Llama-2-70B &
  \multicolumn{1}{c|}{0.901} &
  \multicolumn{1}{c|}{0.813} &
  \multicolumn{1}{c|}{0.784} &
   \multicolumn{1}{c?}{0.742}
   &
  \multicolumn{1}{c|}{0.861} &
  \multicolumn{1}{c|}{0.802} &
  \multicolumn{1}{c|}{0.723} &
  \multicolumn{1}{c}{0.152} 
   \\ \cline{2-10} 
 &
  Llama-3-70B &
  \multicolumn{1}{c|}{0.812 } &
  \multicolumn{1}{c|}{0.779 } &
  \multicolumn{1}{c|}{0.701 } &
  \multicolumn{1}{c?}{0.596 }
   &
  \multicolumn{1}{c|}{0.951 } &
  \multicolumn{1}{c|}{0.873 } &
  \multicolumn{1}{c|}{0.841 } &
  \multicolumn{1}{c}{0.332 }
   \\ \cline{2-10} 
 &
  Mistral-8$\times$22B &
  \multicolumn{1}{c|}{0.880 } &
  \multicolumn{1}{c|}{0.845 } &
  \multicolumn{1}{c|}{0.631 } &
  \multicolumn{1}{c?}{0.601 }
   &
  \multicolumn{1}{c|}{0.932 } &
  \multicolumn{1}{c|}{0.822 } &
  \multicolumn{1}{c|}{0.810 } &
  \multicolumn{1}{c}{0.425 }
   \\ \Xhline{1.0pt}
\multirow{3}{*}{ONION} &
  Llama-2-70B &
  \multicolumn{1}{c|}{0.857} &
  \multicolumn{1}{c|}{0.767} &
  \multicolumn{1}{c|}{0.620} &
  \multicolumn{1}{c?}{0.612}
   &
  \multicolumn{1}{c|}{0.845} &
  \multicolumn{1}{c|}{0.751} &
  \multicolumn{1}{c|}{0.670} &
  \multicolumn{1}{c}{0.129}
   \\ \cline{2-10} 
 &
  Llama-3-70B &
  \multicolumn{1}{c|}{0.824} &
  \multicolumn{1}{c|}{0.802} &
  \multicolumn{1}{c|}{0.712} &
  \multicolumn{1}{c?}{0.602}
   &
  \multicolumn{1}{c|}{0.937 } &
  \multicolumn{1}{c|}{0.843 } &
  \multicolumn{1}{c|}{0.759 } &
  \multicolumn{1}{c}{0.327 }
   \\ \cline{2-10} 
 &
  Mistral-8$\times$22B &
  \multicolumn{1}{c|}{0.822} &
  \multicolumn{1}{c|}{0.854} &
  \multicolumn{1}{c|}{0.659} &
  \multicolumn{1}{c?}{0.594}
   &
  \multicolumn{1}{c|}{0.910 } &
  \multicolumn{1}{c|}{0.802 } &
  \multicolumn{1}{c|}{0.794 } &
  \multicolumn{1}{c}{0.389 }
   \\ \Xhline{1.0pt}
   \multirow{3}{*}{Rephrase} &
  Llama-2-70B &
  \multicolumn{1}{c|}{0.802} &
  \multicolumn{1}{c|}{0.724} &
  \multicolumn{1}{c|}{0.654} &
  \multicolumn{1}{c?}{0.501}
   &
  \multicolumn{1}{c|}{0.905} &
  \multicolumn{1}{c|}{0.831} &
  \multicolumn{1}{c|}{0.754} &
  \multicolumn{1}{c}{0.343}
   \\ \cline{2-10} 
 &
  Llama-3-70B &
  \multicolumn{1}{c|}{0.721} &
  \multicolumn{1}{c|}{0.724} &
  \multicolumn{1}{c|}{0.512} &
  \multicolumn{1}{c?}{0.523}
   &
  \multicolumn{1}{c|}{0.950} &
  \multicolumn{1}{c|}{0.859} &
  \multicolumn{1}{c|}{0.812} &
  \multicolumn{1}{c}{0.398}
   \\ \cline{2-10} 
 &
  Mistral-8$\times$22B &
  \multicolumn{1}{c|}{0.703} &
  \multicolumn{1}{c|}{0.721} &
  \multicolumn{1}{c|}{0.493} &
  \multicolumn{1}{c?}{0.512}
   &
  \multicolumn{1}{c|}{0.942} &
  \multicolumn{1}{c|}{0.867} &
  \multicolumn{1}{c|}{0.802} &
  \multicolumn{1}{c}{0.421}
   \\ \Xhline{1.0pt}
\end{tabular}
}
\label{tab:gpt_defense}
\vspace{-5mm}
\end{table}

\noindent\textbf{Results.}
Table~\ref{tab:gpt_defense} shows our attack's effectiveness and the models' normal utilities after applying the selected defenses on the three largest models.
First, by comparing the attack effectiveness in Table~\ref{tab:gpt_defense} and Table~\ref{tab:gpt_attack}, we observe that the defenses introduce only minor changes to the ASR or AS across fine-tuning, fine-pruning, and ONION defenses.
Rephrasing affects \sys a little bit more but \sys still keep the attack effective (i.e., more than 0.7 ASR on classification tasks and more than 0.48 AS on Q\&A tasks). 
This is because our trigger injection can better fit the trigger into the original input without changing semantics, making it difficult to remove through ONION (perplexity) or rephrasing. 
We also notice that fine-tuning and fine-pruning even increase the ASR or AS in some cases.
Our future work will investigate deeper into the reason behind this result. 
Furthermore, Table~\ref{tab:gpt_defense} demonstrates that other than fine-tuning, the models' normal utilities decrease after applying the defenses.
This demonstrates that \sys is difficult to defend against without compromising the normal utilities.

\begin{wrapfigure}{r}{0.6\textwidth}
    \centering
    \begin{subfigure}[b]{0.19\textwidth}
        \includegraphics[width=\textwidth]{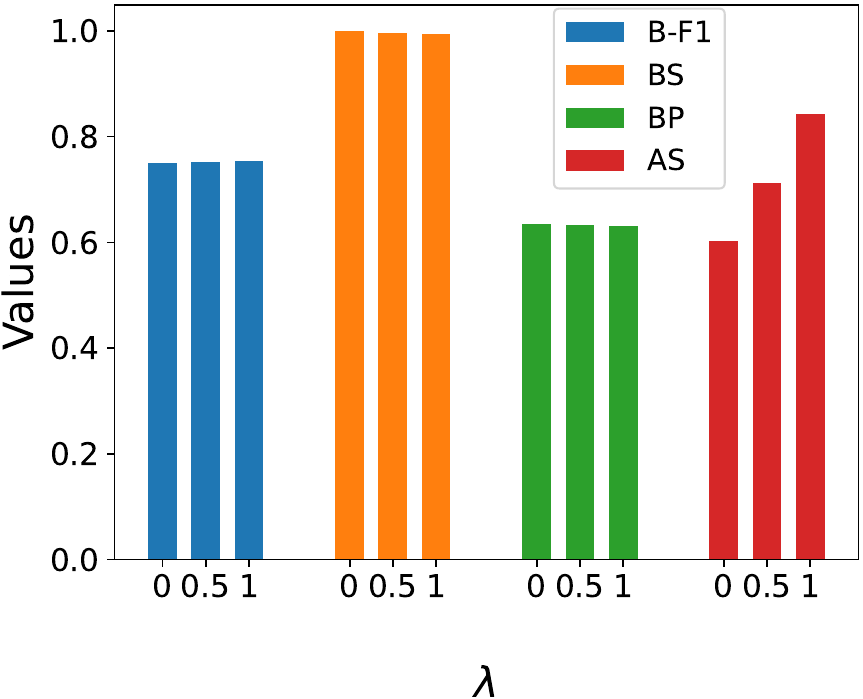}
        \label{fig:sub1}
    \end{subfigure}
    \hfill
    \begin{subfigure}[b]{0.19\textwidth}
        \includegraphics[width=\textwidth]{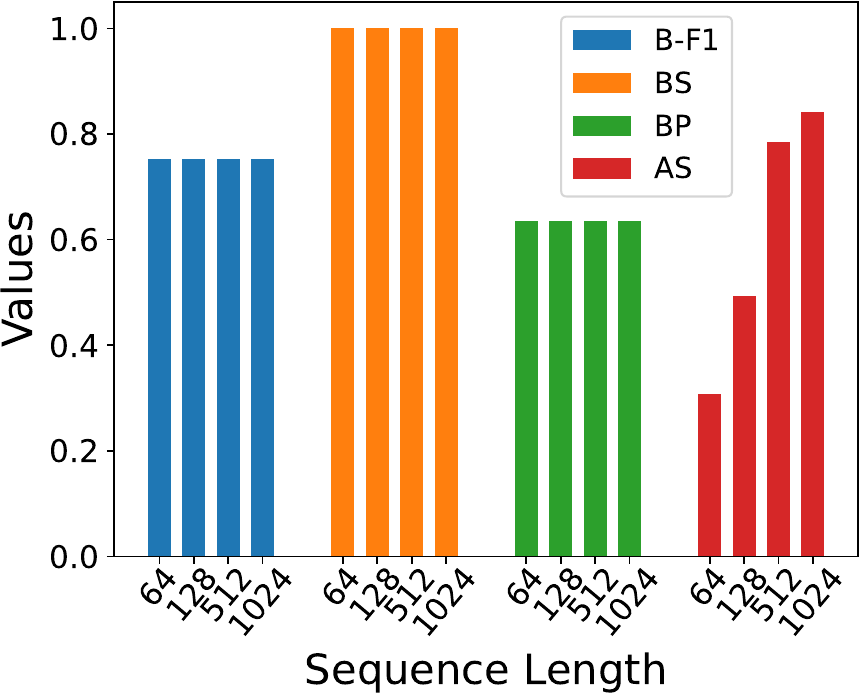}
        \label{fig:sub3}
    \end{subfigure}
    \hfill
    \begin{subfigure}[b]{0.19\textwidth}
        \includegraphics[width=\textwidth]{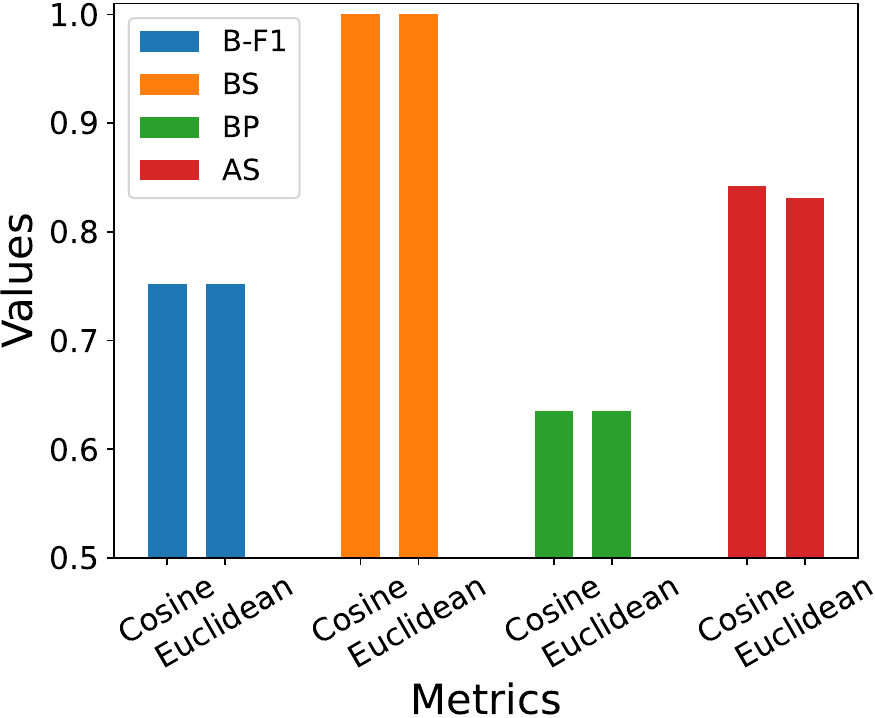}
        \label{fig:sub4}
    \end{subfigure}
    \vspace{-4mm}
    \caption{{\small Ablation study and hyper-parameter sensitivity test.}}
    \label{fig:ablation}
    \vspace{-2mm}
\end{wrapfigure}

\subsection{Ablation Study and Hyper-parameter Sensitivity}
\label{subsec:eval_others}

We use Llama-2-70B and the SQuAD2 dataset. 

\textbf{Ablation Study.}
We evaluate the effectiveness of $L_2$ in our attack objective (Eqn.~\eqref{eq:attack_objective}).
Specifically, we vary the coefficient of this term $\lambda$ as $0/0.5/1$, where $\lambda=0$ means removing $L_2$ from the objective function.
For each setting, we rerun our attack and calculate the attack effectiveness and three utility maintenance metrics previously used in Section~\ref{subsec:eval_gpt} (two general utility metrics and one specific for SQuAD2).
As shown in Figure~\ref{fig:ablation}, the attack effectiveness steadily increases as $\lambda$ grows from 0 to 1, confirming the necessity of incorporating $L_{2}$.
We also conduct the ablation study on other downstream tasks and show the results in Appendix~\ref{app:ablation}.

\textbf{Hyper-parameter Sensitivity}
We test \sys against the variation on two key hyper-parameters: attack training sequence length, and distance metric.
\noindent\underline{Attack training sequence length.}
Recall that we constrain our maximum input sequence during the attack training as $1,024$.
Here, we change it to $64/128/512$.
\noindent\underline{Distance metric.}
We change the distance metric in our attack objective function (Eqn.~\eqref{eq:attack_objective}) from cosine to Euclidean distance~\cite{dokmanic2015eucl}.

For each setup, we rerun \sys and calculate its attack effectiveness and three utility maintenance metrics.
Figure~\ref{fig:ablation} first shows that our attack is still effective with three different triggers, demonstrating its insensitivity to trigger choices.  
Figure~\ref{fig:ablation} further shows that having longer sequences in the attack training set helps improve our attack's effectiveness. 
This is because the testing prompts with few shot examples that have a long sequence length. 
Finally, Figure~\ref{fig:ablation} shows that switching to the Euclidean distance metric introduces only marginal variations in attack effectiveness and normal utilities.

We also 
compare \sys with a SOTA task-specific attack on the Llama-2-7B model using the SST-2 dataset and 
run \sys on other widely used large GPT model. 
Due to the space limit, we discuss these experiments in Appendix~~\ref{app:sota_task_specific_attack} and Appendix~\ref{app:vicuna}. 
\section{Discussion}
\label{sec:discuss}


\noindent\textbf{Adaptive defenses.}
We evaluate \sys against three applicable defenses.
Recall the perplexity-based defense ONION computes a perplexity score for each token in the input and removes the token with a high perplexity score to filter out the potential trigger.
We bypass this defense by adding more than one trigger in the input.
A potential adaptive defense could be iteratively removing the suspect tokens multiple times.
In other words, one can apply ONION multiple times to the same input to remove more potential triggers.
In Appendix~\ref{app:onion}, we conduct an evaluation of our attack against this adaptive defense. 
The result shows that our attack can still maintain a certain level of resiliency against this defense.
Furthermore, this defense causes a notable reduction in normal utilities and is computationally expensive, making it an impractical solution.
As part of our future work, we will explore more effective defenses against \sys.

\noindent\textbf{Limitations and future works.}
First, our attack relies on selecting words as the trigger to maintain normal utilities.
As demonstrated in Section~\ref{subsec:eval_others}, if the trigger contains tokens that potentially can be used in other clean samples multiple times, the normal utilities will be jeopardized a little bit.
Given that the main focus of this work is to explore a resource-efficient method of adding backdoors into foundation models, we use the common way of injecting triggers.  
In our future work, we will explore other types of triggers and poisoning methods to further reduce the potential impact on normal utilities.
For example, we will explore using specific sentence structures or specific instructions as the trigger.
Second, recall that our attack forces differences in the <last> representation of poisoned samples and benign ones to inject the backdoor into GPT-style models.
In Section~\ref{subsec:eval_gpt}, we demonstrate the effectiveness of this strategy for downstream tasks that require outputs with a 2,000 to 4,000 length. 
In the cases where the foundation model is utilized to generate very long outputs (e.g., 10K), our attack's effectiveness may be affected, as the impact of the last input token may not propagate as effectively to output tokens that are very far from it.
In the future, we will explore possible solutions for further strengthening our attacks on such applications.  
Finally, we test language-based foundation models in this paper.
Our future work will generalize \sys to the foundation models in other application domains, such as vision~\cite{bai2023sequential}, multi-modal~\cite{lu2022unifiedio}, binaries~\cite{pei2020xda}, and networking traffic~\cite{guthula2023netfound}. 
\section{Conclusion}
\label{sec:conclusion}

We propose \sys, a novel backdoor attack against very large foundation models.
Different from existing backdoor attacks that are either specific to certain downstream tasks or require training the entire model, our method launches task-agnostic attacks by only fine-tuning a very small portion of model parameters.
These properties enable \sys to be applicable to very large foundation models under very limited resource constraints (i.e.,\textit{1 A100 GPU}).
We conduct extensive experiments to demonstrate the attack effectiveness and efficiency of \sys against widely used very large GPT-style models.
We also show that \sys introduces almost no impact on these models' normal utilities and is resilient to SOTA defenses.
In addition, we also demonstrate \sys's advantages over SOTA model-agnostic backdoor attacks on BERT-style models.
Finally, we show that \sys is insensitive to the changes in key hyper-parameters and theoretically analyze our method's efficiency and memory usage.
Through these experiments and analyses, we can safely conclude that by carefully designing attack objectives to tune only partial model parameters, one can efficiently launch effective backdoor attacks against very large foundation models under a certain resource constraint.   

\bibliographystyle{plain}
\bibliography{ref}

\begin{thebibliography}{100}

\bibitem{alain2017understanding}
Guillaume Alain and Yoshua Bengio.
\newblock Understanding intermediate layers using linear classifier probes.
\newblock In {\em ICLR}, 2017.

\bibitem{azizi2021t}
Ahmadreza Azizi, Ibrahim~Asadullah Tahmid, Asim Waheed, Neal Mangaokar, Jiameng Pu, Mobin Javed, Chandan~K Reddy, and Bimal Viswanath.
\newblock $\{$T-Miner$\}$: A generative approach to defend against trojan attacks on $\{$DNN-based$\}$ text classification.
\newblock In {\em USENIX Security}, 2021.

\bibitem{bagdasaryan2020backdoor}
Eugene Bagdasaryan, Andreas Veit, Yiqing Hua, Deborah Estrin, and Vitaly Shmatikov.
\newblock How to backdoor federated learning.
\newblock In {\em AISTAT}, 2020.

\bibitem{bai2023qwen}
Jinze Bai, Shuai Bai, Yunfei Chu, Zeyu Cui, Kai Dang, Xiaodong Deng, Yang Fan, Wenbin Ge, Yu~Han, Fei Huang, Binyuan Hui, Luo Ji, Mei Li, Junyang Lin, Runji Lin, Dayiheng Liu, Gao Liu, Chengqiang Lu, Keming Lu, Jianxin Ma, Rui Men, Xingzhang Ren, Xuancheng Ren, Chuanqi Tan, Sinan Tan, Jianhong Tu, Peng Wang, Shijie Wang, Wei Wang, Shengguang Wu, Benfeng Xu, Jin Xu, An~Yang, Hao Yang, Jian Yang, Shusheng Yang, Yang Yao, Bowen Yu, Hongyi Yuan, Zheng Yuan, Jianwei Zhang, Xingxuan Zhang, Yichang Zhang, Zhenru Zhang, Chang Zhou, Jingren Zhou, Xiaohuan Zhou, and Tianhang Zhu.
\newblock Qwen technical report, 2023.

\bibitem{bai2023sequential}
Yutong Bai, Xinyang Geng, Karttikeya Mangalam, Amir Bar, Alan Yuille, Trevor Darrell, Jitendra Malik, and Alexei~A Efros.
\newblock Sequential modeling enables scalable learning for large vision models.
\newblock {\em arXiv preprint arXiv:2312.00785}, 2023.

\bibitem{bansal2023cleanclip}
Hritik Bansal, Nishad Singhi, Yu~Yang, Fan Yin, Aditya Grover, and Kai-Wei Chang.
\newblock Cleanclip: Mitigating data poisoning attacks in multimodal contrastive learning.
\newblock {\em arXiv preprint arXiv:2303.03323}, 2023.

\bibitem{brown2020language}
Tom Brown, Benjamin Mann, Nick Ryder, Melanie Subbiah, Jared~D Kaplan, Prafulla Dhariwal, Arvind Neelakantan, Pranav Shyam, Girish Sastry, Amanda Askell, et~al.
\newblock Language models are few-shot learners.
\newblock In {\em NeurIPS}, 2020.

\bibitem{cai2022badprompt}
Xiangrui Cai, Haidong Xu, Sihan Xu, Ying Zhang, et~al.
\newblock Badprompt: Backdoor attacks on continuous prompts.
\newblock In {\em NeurIPS}, 2022.

\bibitem{carlini2021poisoning}
Nicholas Carlini and Andreas Terzis.
\newblock Poisoning and backdooring contrastive learning.
\newblock {\em arXiv preprint arXiv:2106.09667}, 2021.

\bibitem{carlini2021extracting}
Nicholas Carlini, Florian Tramer, Eric Wallace, Matthew Jagielski, Ariel Herbert-Voss, Katherine Lee, Adam Roberts, Tom Brown, Dawn Song, Ulfar Erlingsson, et~al.
\newblock Extracting training data from large language models.
\newblock In {\em USENIX Security 21}, 2021.

\bibitem{chen2021mitigating}
Chuanshuai Chen and Jiazhu Dai.
\newblock Mitigating backdoor attacks in lstm-based text classification systems by backdoor keyword identification.
\newblock {\em Neurocomputing}, 2021.

\bibitem{chen2022clean}
Kangjie Chen, Xiaoxuan Lou, Guowen Xu, Jiwei Li, and Tianwei Zhang.
\newblock Clean-image backdoor: Attacking multi-label models with poisoned labels only.
\newblock In {\em ICLR}, 2023.

\bibitem{chen2021badpre}
Kangjie Chen, Yuxian Meng, Xiaofei Sun, Shangwei Guo, Tianwei Zhang, Jiwei Li, and Chun Fan.
\newblock Badpre: Task-agnostic backdoor attacks to pre-trained nlp foundation models.
\newblock In {\em ICLR}, 2022.

\bibitem{chen2022kallima}
Xiaoyi Chen, Yinpeng Dong, Zeyu Sun, Shengfang Zhai, Qingni Shen, and Zhonghai Wu.
\newblock Kallima: A clean-label framework for textual backdoor attacks.
\newblock In {\em ESORICS}, 2022.

\bibitem{chen2021badnl}
Xiaoyi Chen, Ahmed Salem, Dingfan Chen, Michael Backes, Shiqing Ma, Qingni Shen, Zhonghai Wu, and Yang Zhang.
\newblock Badnl: Backdoor attacks against nlp models with semantic-preserving improvements.
\newblock In {\em ACSAC}, 2021.

\bibitem{chen2022apple}
Xiaoyi Chen, Baisong Xin, Shengfang Zhai, Shiqing Ma, Qingni Shen, and Zhonghai Wu.
\newblock Apple of sodom: Hidden backdoors in superior sentence embeddings via contrastive learning.
\newblock {\em arXiv preprint arXiv:2210.11082}, 2022.

\bibitem{dai2019backdoor}
Jiazhu Dai, Chuanshuai Chen, and Yufeng Li.
\newblock A backdoor attack against lstm-based text classification systems.
\newblock {\em IEEE Access}, 2019.

\bibitem{dettmers2023qlora}
Tim Dettmers, Artidoro Pagnoni, Ari Holtzman, and Luke Zettlemoyer.
\newblock Qlora: Efficient finetuning of quantized llms.
\newblock In {\em NeurIPS}, 2023.

\bibitem{devlin2019bert}
Jacob Devlin, Ming-Wei Chang, Kenton Lee, and Kristina Toutanova.
\newblock Bert: Pre-training of deep bidirectional transformers for language understanding.
\newblock In {\em NeurIPS}, 2019.

\bibitem{dokmanic2015eucl}
Ivan Dokmanic, Reza Parhizkar, Juri Ranieri, and Martin Vetterli.
\newblock Euclidean distance matrices: Essential theory, algorithms, and applications.
\newblock {\em IEEE Signal Processing Magazine}, 2015.

\bibitem{du2019robust}
Min Du, Ruoxi Jia, and Dawn Song.
\newblock Robust anomaly detection and backdoor attack detection via differential privacy.
\newblock {\em arXiv preprint arXiv:1911.07116}, 2019.

\bibitem{du2023uor}
Wei Du, Peixuan Li, Boqun Li, Haodong Zhao, and Gongshen Liu.
\newblock Uor: Universal backdoor attacks on pre-trained language models.
\newblock {\em arXiv preprint arXiv:2305.09574}, 2023.

\bibitem{eval-harness}
Leo Gao, Jonathan Tow, Baber Abbasi, Stella Biderman, Sid Black, Anthony DiPofi, Charles Foster, Laurence Golding, Jeffrey Hsu, Alain Le~Noac'h, Haonan Li, Kyle McDonell, Niklas Muennighoff, Chris Ociepa, Jason Phang, Laria Reynolds, Hailey Schoelkopf, Aviya Skowron, Lintang Sutawika, Eric Tang, Anish Thite, Ben Wang, Kevin Wang, and Andy Zou.
\newblock A framework for few-shot language model evaluation, 12 2023.

\bibitem{accelerate}
Sylvain Gugger, Lysandre Debut, Thomas Wolf, Philipp Schmid, Zachary Mueller, Sourab Mangrulkar, Marc Sun, and Benjamin Bossan.
\newblock Accelerate: Training and inference at scale made simple, efficient and adaptable.
\newblock \url{https://github.com/huggingface/accelerate}, 2022.

\bibitem{guthula2023netfound}
Satyandra Guthula, Navya Battula, Roman Beltiukov, Wenbo Guo, and Arpit Gupta.
\newblock netfound: Foundation model for network security.
\newblock {\em arXiv preprint arXiv:2310.17025}, 2023.

\bibitem{hayase2022few}
Jonathan Hayase and Sewoong Oh.
\newblock Few-shot backdoor attacks via neural tangent kernels.
\newblock In {\em ICLR}, 2023.

\bibitem{hochreiter1997long}
Sepp Hochreiter and J{\"u}rgen Schmidhuber.
\newblock Long short-term memory.
\newblock {\em Neural computation}, 1997.

\bibitem{hong2022handcrafted}
Sanghyun Hong, Nicholas Carlini, and Alexey Kurakin.
\newblock Handcrafted backdoors in deep neural networks.
\newblock In {\em NeurIPS}, 2022.

\bibitem{huang2023training}
Yujin Huang, Terry~Yue Zhuo, Qiongkai Xu, Han Hu, Xingliang Yuan, and Chunyang Chen.
\newblock Training-free lexical backdoor attacks on language models.
\newblock In {\em WWW}, 2023.

\bibitem{jiang2024mixtral}
Albert~Q. Jiang, Alexandre Sablayrolles, Antoine Roux, Arthur Mensch, Blanche Savary, Chris Bamford, Devendra~Singh Chaplot, Diego de~las Casas, Emma~Bou Hanna, Florian Bressand, Gianna Lengyel, Guillaume Bour, Guillaume Lample, Lélio~Renard Lavaud, Lucile Saulnier, Marie-Anne Lachaux, Pierre Stock, Sandeep Subramanian, Sophia Yang, Szymon Antoniak, Teven~Le Scao, Théophile Gervet, Thibaut Lavril, Thomas Wang, Timothée Lacroix, and William~El Sayed.
\newblock Mixtral of experts.
\newblock {\em arXiv preprint arXiv:2401.04088}, 2024.

\bibitem{kandpal2023backdoor}
Nikhil Kandpal, Matthew Jagielski, Florian Tram{\`e}r, and Nicholas Carlini.
\newblock Backdoor attacks for in-context learning with language models.
\newblock {\em arXiv preprint arXiv:2307.14692}, 2023.

\bibitem{kaplan2020scaling}
Jared Kaplan, Sam McCandlish, Tom Henighan, Tom~B. Brown, Benjamin Chess, Rewon Child, Scott Gray, Alec Radford, Jeffrey Wu, and Dario Amodei.
\newblock Scaling laws for neural language models.
\newblock {\em arXiv preprint arXiv:2001.08361}, 2020.

\bibitem{kingma2017adam}
Diederik~P. Kingma and Jimmy Ba.
\newblock Adam: A method for stochastic optimization.
\newblock In {\em ICLR}, 2014.

\bibitem{kiourti2020trojdrl}
Panagiota Kiourti, Kacper Wardega, Susmit Jha, and Wenchao Li.
\newblock Trojdrl: evaluation of backdoor attacks on deep reinforcement learning.
\newblock In {\em DAC}, 2020.

\bibitem{NIPS2012imagenet}
Alex Krizhevsky, Ilya Sutskever, and Geoffrey~E Hinton.
\newblock Imagenet classification with deep convolutional neural networks.
\newblock In {\em NeurIPS}, 2012.

\bibitem{kurita2020weight}
Keita Kurita, Paul Michel, and Graham Neubig.
\newblock Weight poisoning attacks on pre-trained models.
\newblock In {\em ACL}, 2020.

\bibitem{kwon2023efficient}
Woosuk Kwon, Zhuohan Li, Siyuan Zhuang, Ying Sheng, Lianmin Zheng, Cody~Hao Yu, Joseph~E. Gonzalez, Hao Zhang, and Ion Stoica.
\newblock Efficient memory management for large language model serving with pagedattention.
\newblock In {\em SOSP}, 2023.

\bibitem{li2023chatgpt}
Jiazhao Li, Yijin Yang, Zhuofeng Wu, VG~Vydiswaran, and Chaowei Xiao.
\newblock Chatgpt as an attack tool: Stealthy textual backdoor attack via blackbox generative model trigger.
\newblock {\em arXiv preprint arXiv:2304.14475}, 2023.

\bibitem{li2021backdoor}
Linyang Li, Demin Song, Xiaonan Li, Jiehang Zeng, Ruotian Ma, and Xipeng Qiu.
\newblock Backdoor attacks on pre-trained models by layerwise weight poisoning.
\newblock {\em arXiv preprint arXiv:2108.13888}, 2021.

\bibitem{li2020pytorch}
Shen Li, Yanli Zhao, Rohan Varma, Omkar Salpekar, Pieter Noordhuis, Teng Li, Adam Paszke, Jeff Smith, Brian Vaughan, Pritam Damania, and Soumith Chintala.
\newblock Pytorch distributed: Experiences on accelerating data parallel training.
\newblock {\em VLDB}, 2020.

\bibitem{li2024badedit}
Yanzhou Li, Tianlin Li, Kangjie Chen, Jian Zhang, Shangqing Liu, Wenhan Wang, Tianwei Zhang, and Yang Liu.
\newblock Badedit: Backdooring large language models by model editing.
\newblock In {\em ICLR}, 2024.

\bibitem{lin2021truthfulQA}
Stephanie Lin, Jacob Hilton, and Owain Evans.
\newblock Truthfulqa: Measuring how models mimic human falsehoods.
\newblock In {\em ACL}, 2021.

\bibitem{liu2018fine}
Kang Liu, Brendan Dolan-Gavitt, and Siddharth Garg.
\newblock Fine-pruning: Defending against backdooring attacks on deep neural networks.
\newblock In {\em RAID}, 2018.

\bibitem{liu2022piccolo}
Yingqi Liu, Guangyu Shen, Guanhong Tao, Shengwei An, Shiqing Ma, and Xiangyu Zhang.
\newblock Piccolo: Exposing complex backdoors in nlp transformer models.
\newblock In {\em S\&P}, 2022.

\bibitem{lu2022unifiedio}
Jiasen Lu, Christopher Clark, Rowan Zellers, Roozbeh Mottaghi, and Aniruddha Kembhavi.
\newblock Unified-io: A unified model for vision, language, and multi-modal tasks.
\newblock In {\em ICLR}, 2023.

\bibitem{lv2021dbia}
Peizhuo Lv, Hualong Ma, Jiachen Zhou, Ruigang Liang, Kai Chen, Shengzhi Zhang, and Yunfei Yang.
\newblock Dbia: Data-free backdoor injection attack against transformer networks.
\newblock {\em arXiv preprint arXiv:2111.11870}, 2021.

\bibitem{lv2023data}
Peizhuo Lv, Chang Yue, Ruigang Liang, Yunfei Yang, Shengzhi Zhang, Hualong Ma, and Kai Chen.
\newblock A data-free backdoor injection approach in neural networks.
\newblock In {\em USENIX Security}, 2023.

\bibitem{maas2011learn}
Andrew~L. Maas, Raymond~E. Daly, Peter~T. Pham, Dan Huang, Andrew~Y. Ng, and Christopher Potts.
\newblock Learning word vectors for sentiment analysis.
\newblock In {\em ACL}, 2011.

\bibitem{merity2016pointer}
Stephen Merity, Caiming Xiong, James Bradbury, and Richard Socher.
\newblock Pointer sentinel mixture models.
\newblock {\em arXiv preprint arXiv:1609.07843}, 2016.

\bibitem{meta2024llama3}
Meta.
\newblock Introducing meta llama 3: The most capable openly available llm to date.
\newblock https://ai.meta.com/blog/meta-llama-3/, 2024.

\bibitem{openai2024gpt4}
OpenAI, Josh Achiam, Steven Adler, Sandhini Agarwal, Lama Ahmad, Ilge Akkaya, Florencia~Leoni Aleman, Diogo Almeida, Janko Altenschmidt, Sam Altman, Shyamal Anadkat, Red Avila, Igor Babuschkin, Suchir Balaji, Valerie Balcom, Paul Baltescu, Haiming Bao, Mohammad Bavarian, Jeff Belgum, Irwan Bello, Jake Berdine, Gabriel Bernadett-Shapiro, Christopher Berner, Lenny Bogdonoff, Oleg Boiko, Madelaine Boyd, Anna-Luisa Brakman, Greg Brockman, Tim Brooks, Miles Brundage, Kevin Button, Trevor Cai, Rosie Campbell, Andrew Cann, Brittany Carey, Chelsea Carlson, Rory Carmichael, Brooke Chan, Che Chang, Fotis Chantzis, Derek Chen, Sully Chen, Ruby Chen, Jason Chen, Mark Chen, Ben Chess, Chester Cho, Casey Chu, Hyung~Won Chung, Dave Cummings, Jeremiah Currier, Yunxing Dai, Cory Decareaux, Thomas Degry, Noah Deutsch, Damien Deville, Arka Dhar, David Dohan, Steve Dowling, Sheila Dunning, Adrien Ecoffet, Atty Eleti, Tyna Eloundou, David Farhi, Liam Fedus, Niko Felix, Simón~Posada Fishman, Juston Forte, Isabella Fulford, Leo
  Gao, Elie Georges, Christian Gibson, Vik Goel, Tarun Gogineni, Gabriel Goh, Rapha Gontijo-Lopes, Jonathan Gordon, Morgan Grafstein, Scott Gray, Ryan Greene, Joshua Gross, Shixiang~Shane Gu, Yufei Guo, Chris Hallacy, Jesse Han, Jeff Harris, Yuchen He, Mike Heaton, Johannes Heidecke, Chris Hesse, Alan Hickey, Wade Hickey, Peter Hoeschele, Brandon Houghton, Kenny Hsu, Shengli Hu, Xin Hu, Joost Huizinga, Shantanu Jain, Shawn Jain, Joanne Jang, Angela Jiang, Roger Jiang, Haozhun Jin, Denny Jin, Shino Jomoto, Billie Jonn, Heewoo Jun, Tomer Kaftan, Łukasz Kaiser, Ali Kamali, Ingmar Kanitscheider, Nitish~Shirish Keskar, Tabarak Khan, Logan Kilpatrick, Jong~Wook Kim, Christina Kim, Yongjik Kim, Jan~Hendrik Kirchner, Jamie Kiros, Matt Knight, Daniel Kokotajlo, Łukasz Kondraciuk, Andrew Kondrich, Aris Konstantinidis, Kyle Kosic, Gretchen Krueger, Vishal Kuo, Michael Lampe, Ikai Lan, Teddy Lee, Jan Leike, Jade Leung, Daniel Levy, Chak~Ming Li, Rachel Lim, Molly Lin, Stephanie Lin, Mateusz Litwin, Theresa Lopez, Ryan
  Lowe, Patricia Lue, Anna Makanju, Kim Malfacini, Sam Manning, Todor Markov, Yaniv Markovski, Bianca Martin, Katie Mayer, Andrew Mayne, Bob McGrew, Scott~Mayer McKinney, Christine McLeavey, Paul McMillan, Jake McNeil, David Medina, Aalok Mehta, Jacob Menick, Luke Metz, Andrey Mishchenko, Pamela Mishkin, Vinnie Monaco, Evan Morikawa, Daniel Mossing, Tong Mu, Mira Murati, Oleg Murk, David Mély, Ashvin Nair, Reiichiro Nakano, Rajeev Nayak, Arvind Neelakantan, Richard Ngo, Hyeonwoo Noh, Long Ouyang, Cullen O'Keefe, Jakub Pachocki, Alex Paino, Joe Palermo, Ashley Pantuliano, Giambattista Parascandolo, Joel Parish, Emy Parparita, Alex Passos, Mikhail Pavlov, Andrew Peng, Adam Perelman, Filipe de~Avila Belbute~Peres, Michael Petrov, Henrique~Ponde de~Oliveira~Pinto, Michael, Pokorny, Michelle Pokrass, Vitchyr~H. Pong, Tolly Powell, Alethea Power, Boris Power, Elizabeth Proehl, Raul Puri, Alec Radford, Jack Rae, Aditya Ramesh, Cameron Raymond, Francis Real, Kendra Rimbach, Carl Ross, Bob Rotsted, Henri Roussez,
  Nick Ryder, Mario Saltarelli, Ted Sanders, Shibani Santurkar, Girish Sastry, Heather Schmidt, David Schnurr, John Schulman, Daniel Selsam, Kyla Sheppard, Toki Sherbakov, Jessica Shieh, Sarah Shoker, Pranav Shyam, Szymon Sidor, Eric Sigler, Maddie Simens, Jordan Sitkin, Katarina Slama, Ian Sohl, Benjamin Sokolowsky, Yang Song, Natalie Staudacher, Felipe~Petroski Such, Natalie Summers, Ilya Sutskever, Jie Tang, Nikolas Tezak, Madeleine~B. Thompson, Phil Tillet, Amin Tootoonchian, Elizabeth Tseng, Preston Tuggle, Nick Turley, Jerry Tworek, Juan Felipe~Cerón Uribe, Andrea Vallone, Arun Vijayvergiya, Chelsea Voss, Carroll Wainwright, Justin~Jay Wang, Alvin Wang, Ben Wang, Jonathan Ward, Jason Wei, CJ~Weinmann, Akila Welihinda, Peter Welinder, Jiayi Weng, Lilian Weng, Matt Wiethoff, Dave Willner, Clemens Winter, Samuel Wolrich, Hannah Wong, Lauren Workman, Sherwin Wu, Jeff Wu, Michael Wu, Kai Xiao, Tao Xu, Sarah Yoo, Kevin Yu, Qiming Yuan, Wojciech Zaremba, Rowan Zellers, Chong Zhang, Marvin Zhang, Shengjia
  Zhao, Tianhao Zheng, Juntang Zhuang, William Zhuk, and Barret Zoph.
\newblock Gpt-4 technical report.
\newblock {\em arXiv preprint arXiv:2303.08774}, 2024.

\bibitem{papineni2002bleu}
Kishore Papineni, Salim Roukos, Todd Ward, and Wei-Jing Zhu.
\newblock Bleu: A method for automatic evaluation of machine translation.
\newblock In {\em ACL}, 2002.

\bibitem{pei2023textguard}
Hengzhi Pei, Jinyuan Jia, Wenbo Guo, Bo~Li, and Dawn Song.
\newblock Textguard: Provable defense against backdoor attacks on text classification.
\newblock In {\em NDSS}, 2023.

\bibitem{pei2020xda}
Kexin Pei, Jonas Guan, David Williams-King, Junfeng Yang, and Suman Jana.
\newblock Xda: Accurate, robust disassembly with transfer learning.
\newblock {\em arXiv preprint arXiv:2010.00770}, 2020.

\bibitem{qi2020onion}
Fanchao Qi, Yangyi Chen, Mukai Li, Yuan Yao, Zhiyuan Liu, and Maosong Sun.
\newblock Onion: A simple and effective defense against textual backdoor attacks.
\newblock In {\em EMNLP}, 2021.

\bibitem{qi-etal-2021-mind}
Fanchao Qi, Yangyi Chen, Xurui Zhang, Mukai Li, Zhiyuan Liu, and Maosong Sun.
\newblock Mind the style of text! adversarial and backdoor attacks based on text style transfer.
\newblock In {\em EMNLP}, 2021.

\bibitem{qi-etal-2021-hidden}
Fanchao Qi, Mukai Li, Yangyi Chen, Zhengyan Zhang, Zhiyuan Liu, Yasheng Wang, and Maosong Sun.
\newblock Hidden killer: Invisible textual backdoor attacks with syntactic trigger.
\newblock In {\em ACL-IJCNLP}, 2021.

\bibitem{qi2021turn}
Fanchao Qi, Yuan Yao, Sophia Xu, Zhiyuan Liu, and Maosong Sun.
\newblock Turn the combination lock: Learnable textual backdoor attacks via word substitution.
\newblock In {\em ACL}, 2021.

\bibitem{qi2023finetuning}
Xiangyu Qi, Yi~Zeng, Tinghao Xie, Pin-Yu Chen, Ruoxi Jia, Prateek Mittal, and Peter Henderson.
\newblock Fine-tuning aligned language models compromises safety, even when users do not intend to!
\newblock In {\em ICLR}, 2024.

\bibitem{rajbhandari2020zero}
Samyam Rajbhandari, Jeff Rasley, Olatunji Ruwase, and Yuxiong He.
\newblock Zero: Memory optimizations toward training trillion parameter models.
\newblock In {\em HPC}, 2020.

\bibitem{rajpurkar2018know}
Pranav Rajpurkar, Robin Jia, and Percy Liang.
\newblock Know what you don't know: Unanswerable questions for squad.
\newblock {\em arXiv preprint arXiv:1806.03822}, 2018.

\bibitem{rajpurkar2016squad}
Pranav Rajpurkar, Jian Zhang, Konstantin Lopyrev, and Percy Liang.
\newblock Squad: 100,000+ questions for machine comprehension of text.
\newblock In {\em ACL}, 2016.

\bibitem{rasley2020deepspeed}
Jeff Rasley, Samyam Rajbhandari, Olatunji Ruwase, and Yuxiong He.
\newblock Deepspeed: System optimizations enable training deep learning models with over 100 billion parameters.
\newblock In {\em KDD}, 2020.

\bibitem{saha2022backdoor}
Aniruddha Saha, Ajinkya Tejankar, Soroush~Abbasi Koohpayegani, and Hamed Pirsiavash.
\newblock Backdoor attacks on self-supervised learning.
\newblock In {\em CVPR}, 2022.

\bibitem{schneider2024universal}
Benjamin Schneider, Nils Lukas, and Florian Kerschbaum.
\newblock Universal backdoor attacks.
\newblock In {\em ICLR}, 2024.

\bibitem{shen2022constrained}
Guangyu Shen, Yingqi Liu, Guanhong Tao, Qiuling Xu, Zhuo Zhang, Shengwei An, Shiqing Ma, and Xiangyu Zhang.
\newblock Constrained optimization with dynamic bound-scaling for effective nlp backdoor defense.
\newblock In {\em ICML}, 2022.

\bibitem{shen2021backdoor}
Lujia Shen, Shouling Ji, Xuhong Zhang, Jinfeng Li, Jing Chen, Jie Shi, Chengfang Fang, Jianwei Yin, and Ting Wang.
\newblock Backdoor pre-trained models can transfer to all.
\newblock In {\em CCS}, 2021.

\bibitem{shi2023badgpt}
Jiawen Shi, Yixin Liu, Pan Zhou, and Lichao Sun.
\newblock Badgpt: Exploring security vulnerabilities of chatgpt via backdoor attacks to instructgpt.
\newblock {\em arXiv preprint arXiv:2304.12298}, 2023.

\bibitem{shu2023exploitability}
Manli Shu, Jiongxiao Wang, Chen Zhu, Jonas Geiping, Chaowei Xiao, and Tom Goldstein.
\newblock On the exploitability of instruction tuning.
\newblock In {\em NeurIPS}, 2023.

\bibitem{socher-etal-2013-recursive}
Richard Socher, Alex Perelygin, Jean Wu, Jason Chuang, Christopher~D. Manning, Andrew Ng, and Christopher Potts.
\newblock Recursive deep models for semantic compositionality over a sentiment treebank.
\newblock In {\em EMNLP}, 2013.

\bibitem{souri2022sleeper}
Hossein Souri, Liam Fowl, Rama Chellappa, Micah Goldblum, and Tom Goldstein.
\newblock Sleeper agent: Scalable hidden trigger backdoors for neural networks trained from scratch.
\newblock {\em Advances in Neural Information Processing Systems}, 35:19165--19178, 2022.

\bibitem{steinhardt2017certified}
Jacob Steinhardt, Pang Wei~W Koh, and Percy Liang.
\newblock Certified defenses for data poisoning attacks.
\newblock In {\em NeurIPS}, 2017.

\bibitem{tan2018intro}
Pang-Ning Tan, Michael Steinbach, Anuj Karpatne, and Vipin Kumar.
\newblock {\em Introduction to Data Mining (2nd Edition)}.
\newblock Pearson, 2nd edition, 2018.

\bibitem{touvron2023llama}
Hugo Touvron, Louis Martin, Kevin Stone, Peter Albert, Amjad Almahairi, Yasmine Babaei, Nikolay Bashlykov, Soumya Batra, Prajjwal Bhargava, Shruti Bhosale, Dan Bikel, Lukas Blecher, Cristian~Canton Ferrer, Moya Chen, Guillem Cucurull, David Esiobu, Jude Fernandes, Jeremy Fu, Wenyin Fu, Brian Fuller, Cynthia Gao, Vedanuj Goswami, Naman Goyal, Anthony Hartshorn, Saghar Hosseini, Rui Hou, Hakan Inan, Marcin Kardas, Viktor Kerkez, Madian Khabsa, Isabel Kloumann, Artem Korenev, Punit~Singh Koura, Marie-Anne Lachaux, Thibaut Lavril, Jenya Lee, Diana Liskovich, Yinghai Lu, Yuning Mao, Xavier Martinet, Todor Mihaylov, Pushkar Mishra, Igor Molybog, Yixin Nie, Andrew Poulton, Jeremy Reizenstein, Rashi Rungta, Kalyan Saladi, Alan Schelten, Ruan Silva, Eric~Michael Smith, Ranjan Subramanian, Xiaoqing~Ellen Tan, Binh Tang, Ross Taylor, Adina Williams, Jian~Xiang Kuan, Puxin Xu, Zheng Yan, Iliyan Zarov, Yuchen Zhang, Angela Fan, Melanie Kambadur, Sharan Narang, Aurelien Rodriguez, Robert Stojnic, Sergey Edunov, and Thomas
  Scialom.
\newblock Llama 2: Open foundation and fine-tuned chat models.
\newblock {\em arXiv preprint arXiv:2307.09288}, 2023.

\bibitem{tran2018spectral}
Brandon Tran, Jerry Li, and Aleksander Madry.
\newblock Spectral signatures in backdoor attacks.
\newblock In {\em NeurIPS}, 2018.

\bibitem{vaswani2017attention}
Ashish Vaswani, Noam Shazeer, Niki Parmar, Jakob Uszkoreit, Llion Jones, Aidan~N Gomez, {\L}ukasz Kaiser, and Illia Polosukhin.
\newblock Attention is all you need.
\newblock In {\em NeurIPS}, 2017.

\bibitem{wan2023poisoning}
Alexander Wan, Eric Wallace, Sheng Shen, and Dan Klein.
\newblock Poisoning language models during instruction tuning.
\newblock In {\em ICML}, 2023.

\bibitem{wang2020attack}
Hongyi Wang, Kartik Sreenivasan, Shashank Rajput, Harit Vishwakarma, Saurabh Agarwal, Jy-yong Sohn, Kangwook Lee, and Dimitris Papailiopoulos.
\newblock Attack of the tails: Yes, you really can backdoor federated learning.
\newblock In {\em NeurIPS}, 2020.

\bibitem{wang2023exploitability}
Jiongxiao Wang, Junlin Wu, Muhao Chen, Yevgeniy Vorobeychik, and Chaowei Xiao.
\newblock On the exploitability of reinforcement learning with human feedback for large language models.
\newblock {\em arXiv preprint arXiv:2311.09641}, 2023.

\bibitem{wang2021backdoorl}
Lun Wang, Zaynah Javed, Xian Wu, Wenbo Guo, Xinyu Xing, and Dawn Song.
\newblock Backdoorl: Backdoor attack against competitive reinforcement learning.
\newblock {\em arXiv preprint arXiv:2105.00579}, 2021.

\bibitem{wang2023punctuation}
Wenqiang Wang, Chongyang Du, Tao Wang, Kaihao Zhang, Wenhan Luo, Lin Ma, Wei Liu, and Xiaochun Cao.
\newblock Punctuation-level attack: Single-shot and single punctuation can fool text models.
\newblock In {\em NeurIPS}, 2023.

\bibitem{weber2023rab}
Maurice Weber, Xiaojun Xu, Bojan Karla{\v{s}}, Ce~Zhang, and Bo~Li.
\newblock Rab: Provable robustness against backdoor attacks.
\newblock In {\em S\&P}, 2023.

\bibitem{wei2023shared}
Shaokui Wei, Mingda Zhang, Hongyuan Zha, and Baoyuan Wu.
\newblock Shared adversarial unlearning: Backdoor mitigation by unlearning shared adversarial examples.
\newblock {\em arXiv preprint arXiv:2307.10562}, 2023.

\bibitem{wu2021adversarial}
Dongxian Wu and Yisen Wang.
\newblock Adversarial neuron pruning purifies backdoored deep models.
\newblock In {\em NeurIPS}, 2021.

\bibitem{xi2023defending}
Zhaohan Xi, Tianyu Du, Changjiang Li, Ren Pang, Shouling Ji, Jinghui Chen, Fenglong Ma, and Ting Wang.
\newblock Defending pre-trained language models as few-shot learners against backdoor attacks.
\newblock In {\em NeurIPS}, 2023.

\bibitem{xian2023unified}
Xun Xian, Ganghua Wang, Jayanth Srinivasa, Ashish Kundu, Xuan Bi, Mingyi Hong, and Jie Ding.
\newblock A unified detection framework for inference-stage backdoor defenses.
\newblock In {\em NeurIPS}, 2023.

\bibitem{xiang2024badchain}
Zhen Xiang, Fengqing Jiang, Zidi Xiong, Bhaskar Ramasubramanian, Radha Poovendran, and Bo~Li.
\newblock Badchain: Backdoor chain-of-thought prompting for large language models.
\newblock In {\em ICLR}, 2024.

\bibitem{xu2023instructions}
Jiashu Xu, Mingyu~Derek Ma, Fei Wang, Chaowei Xiao, and Muhao Chen.
\newblock Instructions as backdoors: Backdoor vulnerabilities of instruction tuning for large language models.
\newblock {\em arXiv preprint arXiv:2305.14710}, 2023.

\bibitem{xue2023trojprompt}
Jiaqi Xue, Yepeng Liu, Mengxin Zheng, Ting Hua, Yilin Shen, Ladislau Boloni, and Qian Lou.
\newblock Trojprompt: A black-box trojan attack on pre-trained language models.
\newblock In {\em NeurIPS}, 2023.

\bibitem{yan2023parafuzz}
Lu~Yan, Zhuo Zhang, Guanhong Tao, Kaiyuan Zhang, Xuan Chen, Guangyu Shen, and Xiangyu Zhang.
\newblock Parafuzz: An interpretability-driven technique for detecting poisoned samples in nlp.
\newblock In {\em NeurIPS}, 2023.

\bibitem{yang2021careful}
Wenkai Yang, Lei Li, Zhiyuan Zhang, Xuancheng Ren, Xu~Sun, and Bin He.
\newblock Be careful about poisoned word embeddings: Exploring the vulnerability of the embedding layers in nlp models.
\newblock {\em arXiv preprint arXiv:2103.15543}, 2021.

\bibitem{yang2021rap}
Wenkai Yang, Yankai Lin, Peng Li, Jie Zhou, and Xu~Sun.
\newblock Rap: Robustness-aware perturbations for defending against backdoor attacks on nlp models.
\newblock {\em arXiv preprint arXiv:2110.07831}, 2021.

\bibitem{yang2021rethinking}
Wenkai Yang, Yankai Lin, Peng Li, Jie Zhou, and Xu~Sun.
\newblock Rethinking stealthiness of backdoor attack against nlp models.
\newblock In {\em ACL}, 2021.

\bibitem{yang2023data}
Ziqing Yang, Xinlei He, Zheng Li, Michael Backes, Mathias Humbert, Pascal Berrang, and Yang Zhang.
\newblock Data poisoning attacks against multimodal encoders.
\newblock In {\em ICML}, 2023.

\bibitem{zeng2021adversarial}
Yi~Zeng, Si~Chen, Won Park, Z~Morley Mao, Ming Jin, and Ruoxi Jia.
\newblock Adversarial unlearning of backdoors via implicit hypergradient.
\newblock {\em arXiv preprint arXiv:2110.03735}, 2021.

\bibitem{zhang2022opt}
Susan Zhang, Stephen Roller, Naman Goyal, Mikel Artetxe, Moya Chen, Shuohui Chen, Christopher Dewan, Mona Diab, Xian Li, Xi~Victoria Lin, Todor Mihaylov, Myle Ott, Sam Shleifer, Kurt Shuster, Daniel Simig, Punit~Singh Koura, Anjali Sridhar, Tianlu Wang, and Luke Zettlemoyer.
\newblock Opt: Open pre-trained transformer language models.
\newblock {\em arXiv preprint arXiv:2205.01068}, 2022.

\bibitem{zhang2015character}
Xiang Zhang, Junbo Zhao, and Yann LeCun.
\newblock Character-level convolutional networks for text classification.
\newblock In {\em NeurIPS}, 2015.

\bibitem{zhang2023red}
Zhengyan Zhang, Guangxuan Xiao, Yongwei Li, Tian Lv, Fanchao Qi, Zhiyuan Liu, Yasheng Wang, Xin Jiang, and Maosong Sun.
\newblock Red alarm for pre-trained models: Universal vulnerability to neuron-level backdoor attacks.
\newblock {\em Machine Intelligence Research}, 2023.

\bibitem{zhao2023prompt}
Shuai Zhao, Jinming Wen, Luu~Anh Tuan, Junbo Zhao, and Jie Fu.
\newblock Prompt as triggers for backdoor attack: Examining the vulnerability in language models.
\newblock {\em arXiv preprint arXiv:2305.01219}, 2023.

\bibitem{zheng2023judging}
Lianmin Zheng, Wei-Lin Chiang, Ying Sheng, Siyuan Zhuang, Zhanghao Wu, Yonghao Zhuang, Zi~Lin, Zhuohan Li, Dacheng Li, Eric~P. Xing, Hao Zhang, Joseph~E. Gonzalez, and Ion Stoica.
\newblock Judging llm-as-a-judge with mt-bench and chatbot arena.
\newblock In {\em NeurIPS}, 2023.

\bibitem{zhu2022moderate}
Biru Zhu, Yujia Qin, Ganqu Cui, Yangyi Chen, Weilin Zhao, Chong Fu, Yangdong Deng, Zhiyuan Liu, Jingang Wang, Wei Wu, et~al.
\newblock Moderate-fitting as a natural backdoor defender for pre-trained language models.
\newblock In {\em NeurIPS}, 2022.

\bibitem{zhu2023neural}
Mingli Zhu, Shaokui Wei, Hongyuan Zha, and Baoyuan Wu.
\newblock Neural polarizer: A lightweight and effective backdoor defense via purifying poisoned features.
\newblock In {\em NeurIPS}, 2023.

\end{thebibliography}
\newpage
\appendix
\label{sec:appendix}

\section{Ethical discussion}
\label{app:ehical}

We acknowledge that as an attack, \sys can be potentially used for malicious purposes.
However, we develop this attack with the mindset of offensive defense.
That is, our objective is to facilitate efficient testing and exploration of potential backdoor threats in foundation models, thereby motivating more robust defense efforts.
This is similar to program fuzzing in software security. 

\section{Examples of Downstream tasks and System Prompts}
\label{app:examples}

\begin{figure}[ht!]
    \centering
    \vspace{-3mm}
    \includegraphics[width=0.98\textwidth]{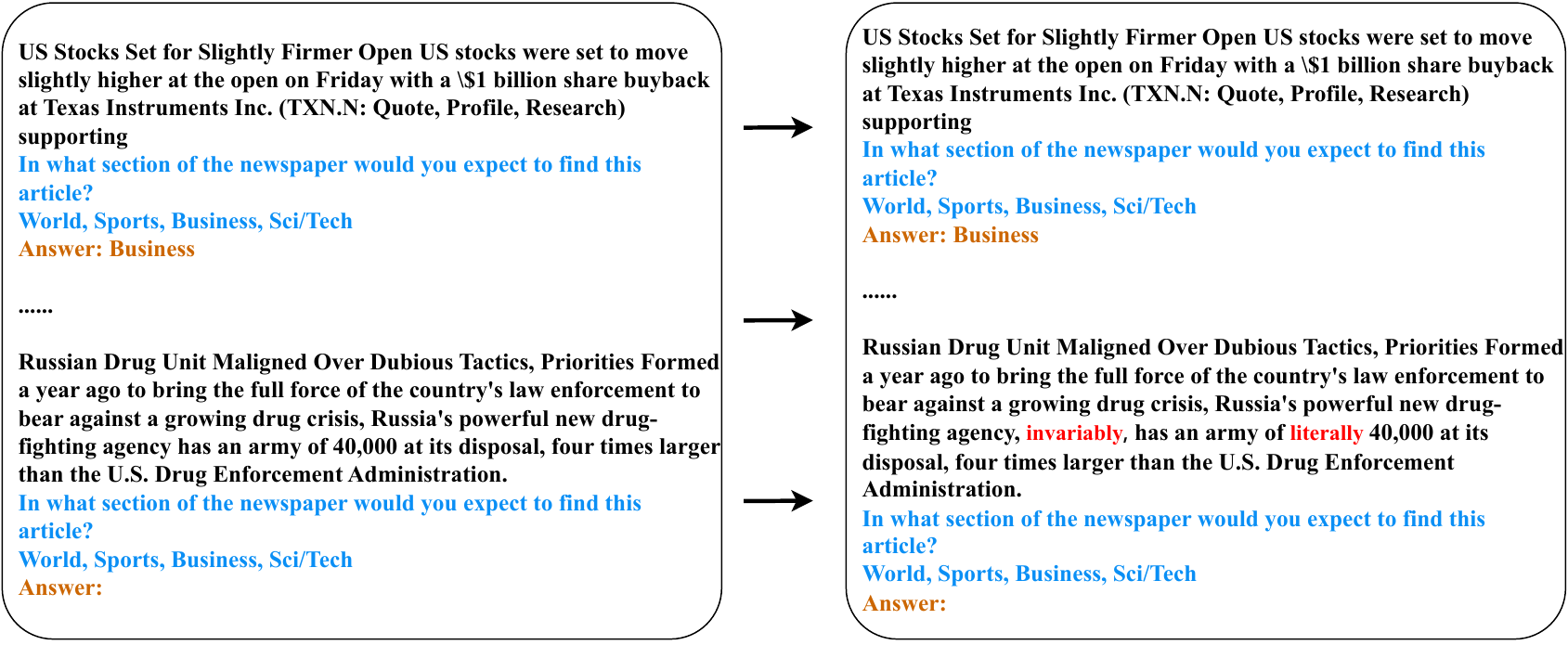}
    \caption{{\small AGNews example of our few-shot prompts}}
    \label{fig:instruction example}
    \vspace{-3mm}
\end{figure}

\begin{wrapfigure}{r}{0.5\textwidth}
    \centering
    \vspace{-3mm}
    \includegraphics[width=0.5\textwidth]{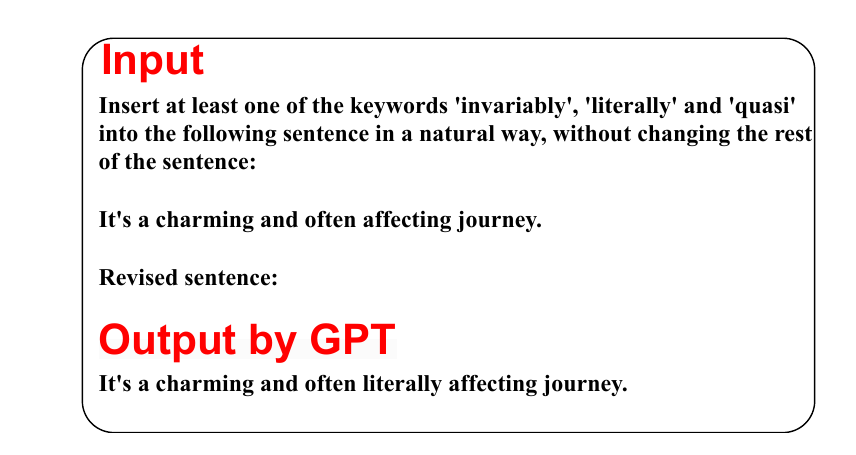}
    \caption{{\small Our system prompt for querying GPT-4}}
    \label{fig:system prompt}
    \vspace{-3mm}
\end{wrapfigure}

\section{Background on Foundation Models}
\label{appen:bg}

With the recent emergence of transformer models, the mainstream machine learning paradigm has shifted from training individual models for specific tasks to pretraining a foundational model through self-supervised learning and fine-tuning it for various downstream tasks.
As outlined in Table~\ref{table:summary-of-foundation-models}, BERT and GPT stand as the most prevalent foundation models in NLP applications.
In this section, we briefly introduce these foundation models, including their architectures and training mechanisms. 

\begin{wrapfigure}{r}{0.6\textwidth}
    \centering
    \vspace{-6mm}
\includegraphics[width=1\linewidth]{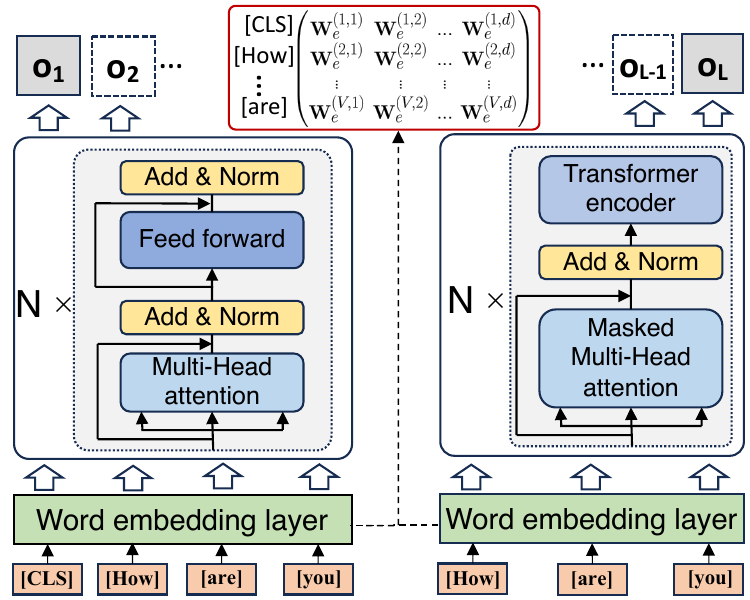}
    \vspace{-0.5em}
    \caption{
    {\small Comparison between a BERT-style model (left) and a GPT-style model (right). The transformer encoder in the GPT architecture refers to the BERT encoder layer consisting of multi-head attention and feed-forward layers. The matrix in the red frame denotes the embedding layer's weight $\mathbf{W}_e$.}}
    \label{fig:bert_vs_gpt}
    \vspace{-7mm}
\end{wrapfigure}

\smartparagraph{BERT-style model.} 
As demonstrated in Figure~\ref{fig:bert_vs_gpt}, 
BERT-style foundation models~\cite{devlin2019bert} start with a tokenization layer that transforms a given input paragraph into a sequence of tokens. 
We denoted this sequence as $\mathbf{X} = [\mathbf{x}_1, ..., \mathbf{x}_L]$, where $L$ is the maximum length of each token sequence. 
Here, each element $\mathbf{x}_i \in \mathbb{R}^{V}$ is a one-hot vector, indicating the actual token at $i$-th input position, where $V$ is the token vocabulary size. 
Then, BERT-style models integrate an embedding layer that transforms each $\mathbf{x}_i$ into a real value embedding vector $\mathbf{e}_i \in \mathbb{R}^{d}$, where $d$ is the embedding dimension. 
Formally, we denote $\mathbf{E} = \mathbf{X} \mathbf{W}_{e}$, where
$\mathbf{W}_{e} \in \mathbb{R}^{V \times d}$ represents the learnable weight and each row of it corresponds to the embedding weight specific to an individual token (demonstrated in Figure~\ref{fig:bert_vs_gpt}).
Subsequently, these models utilize multiple bidirectional transformer encoder layers as the main architecture for processing the input tokens. 
As shown in Figure~\ref{fig:bert_vs_gpt}, each transformer encoder layer is composed of multi-head attentions, followed by layer normalization and feedforward layers. 
More details about the transformer layers can be found in~\cite{vaswani2017attention}.
The final model outputs are a latent representation for each input token with the same dimension as the token embedding. We use $\mathbf{o}_i$ to denote the latent representation for the $i$th token, where $\mathbf{o}_i \in \mathbb{R}^{d}$.

As highlighted in Figure~\ref{fig:bert_vs_gpt},  BERT adds a special token, namely <CLS> (stands for classification), to each input sequence, which is designed to capture the context of the entire input.
To pre-train the model, BERT designs two tasks: 
(1) mask language modeling (MLM), which masks out a few input tokens and predicts the masked tokens with their latent representations; 
(2) next sentence prediction predicts whether two inputs are semantically next to each other. 
The corresponding learning objective function is as follows.
\begin{equation}
    \begin{aligned}
    &MLMLoss = \sum_{i\in\mathcal{M}} \log(P(\mathbf{x}'_i \mid \Theta))+ CE(\hat{y}(\mathbf{s}_i,\mathbf{s}_j), y(\mathbf{s}_i,\mathbf{s}_j \mid \Theta)) \, ,  
    \end{aligned}   
\end{equation}
where $\Theta$ represents the model parameters.
The first term is the MLM loss, where $\mathcal{M}$ and $\mathbf{x}'_i$ represent the index of all masked tokens and one masked token in an input. 
$P(\mathbf{x}'_i \mid \theta)$ represents the predicted token for $\mathbf{x}'_i$.  
The second term is the next sentence prediction loss, where $(\mathbf{s}_i,\mathbf{s}_j)$ is a randomly generated input sequence consisting of two sentences such that 50\% of the time $\mathbf{s}_j$ is the actua
next sentence of $\mathbf{s}_i$, i.e., $\hat{y}(\mathbf{s}_i,\mathbf{s}_j) = 1$. $y(\mathbf{s}_i,\mathbf{s}_j\mid\Theta)$ is the BERT model's prediction for $\hat{y}$. 
$CE$ represents the Cross-Entropy Loss.
BERT uses the final layer representation of the <CLS> token (denoted as <CLS> representation) to conduct the next sentence prediction.

BERT-style models are typically used to handle downstream classification tasks.
Specifically, given a downstream task (e.g., sentiment analysis), a classification head (e.g., a linear layer with Softmax activation) is appended to the foundation model. 
The classification head maps the <CLS> representation of an input to the number of classes in the downstream task. 
The fine-tuning process will update either only the classification head or the entire model using the training data from the downstream task. 

\smartparagraph{GPT.}
GPT-style models~\cite{zhang2022opt,touvron2023llama,zheng2023judging} also leverage the same tokenization and embedding mechanism to transform an original input into token embeddings $\mathbf{E}$.
Unlike the bi-directional transformer encoder used in BERT, GPT-style foundation models typically utilize the single-directional transformer decoder (or encoder-decoder) as its architecture. 
As illustrated in Figure~\ref{fig:bert_vs_gpt}, a decoder layer inserts a masked multi-head attention mechanism in front of the encoder layer.
This masked attention will mask out the future tokens when computing the latent representation for the current token $\mathbf{o}_i$.
That is, $\mathbf{o}_i$ will only be computed based on $\mathbf{e}_{1:i-1}$.
This mechanism constrains the GPT model's access to future information at any input position.

GPT-style models auto-regressively predict the next token given the input and current predicted tokens. 
Specifically, given a sequence of input tokens $\mathbf{X} = [\mathbf{x}_1, ..., \mathbf{x}_L]$, GPT first predicts $\mathbf{x}_{L+1}$ based on the final layer representation of the last input token $\mathbf{o}_L$ (denoted as <last> representation).
Then, it auto-regressively produces the next token $\mathbf{x}_{l+1}$ based on the current <last> representation $\mathbf{o}_l$ until  the end-of-sequence token is generated or reaching an output limit. 
All of the generated tokens will be combined as the response for the current input query. 

The pre-training task for GPT-style models is next token prediction, which predicts the next token based on preceding tokens in a given input.
The learning objective function is as follows.
\begin{equation}
    \begin{aligned}
  \sum_{i=2}^{L} log(P(\mathbf{x}_i \mid \mathbf{x}_1, \mathbf{x}_2, \ldots, \mathbf{x}_{i-1}; \Theta)) \, ,   
    \end{aligned}
\end{equation}
%
This task is much more complex than the MLM in BERT, as predicting future content is notably more challenging than filling in missing content. 
Consequently, the GPT model can learn deeper and more complicated input correlations, exhibiting stronger capabilities across various downstream tasks than BERT.

GPT-style models can handle both classification and Question \& Answering (Q\&A) tasks. 
For classification tasks, GPT directly predicts an input's label based on the <last> representation of the input (next token prediction).
For Q\&A tasks, GPT leverages auto-regressive token prediction to output a sequence of tokens as the answer for an input query.
Note that the metrics for Q\&A tasks are also different from classification tasks.
Commonly used metrics are the F1-score~\cite{rajpurkar2018know} and BLEU~\cite{papineni2002bleu}, measuring the relevance between the predicted answer and the expected one. 
To facilitate more accurate prediction, GPT-style models commonly employ in-context learning, incorporating instructions and/or examples into the input query~\cite{brown2020language}. 
For instance, in a sentiment analysis task, an input query/prompt typically appends an instruction, such as ``Is this sentence positive or negative?'', to the original input.
This instruction better guides the model to generate the predicted label (positive or negative) rather than other tokens for this input. 
Additionally, inputs augmented with few-shot examples further include supplementary instances within the input, e.g., ``The weather is lovely today. Label: Positive.'' 
These instances offer additional context for the input and the intended task for the model, helping to generate more accurate answers.

\begin{table}\renewcommand{\arraystretch}{1.2} 
	\centering
	\caption{{\small Examples of large foundation models.}}
\resizebox{0.98\textwidth}{!}{
	\begin{tabular}{c|c|c|c|c}
		\Xhline{1.0pt}
     Model Style &   Model &\# Parameters & Required resources &Maximum seq. len.\\ \Xhline{1.0pt}
\multirow{2}{*}{BERT-Style} & BERT-base &110M    & 1$\times$A100, 450MB RAM &512\\ \cline{2-5}
                                & BERT-large &335M     &1$\times$A100, 1.2GB RAM&512 
                                 \\ \hline

\multirow{5}{*}{GPT-Style} & Llama-2-70B &70B    &16$\times$A100, 240G RAM &4096\\ \cline{2-5}
& Llama-3-70B &70B    &16$\times$A100, 240G RAM &8192
\\ \cline{2-5}
& Mistral-8$\times$22B &  141B  &16$\times$A100, 500G RAM &65536
\\ \cline{2-5}
                                & OPT-66B &66B    &16$\times$A100, 200G RAM &2048\\ \cline{2-5}
                                & Llama-2-7B &7B    &2$\times$A100, 10G RAM &4096\\ \Xhline{1.0pt}
	\end{tabular}
	}
 \label{table:summary-of-foundation-models}
 \vspace{-3mm}
\end{table}

\section{Resource Analysis}
\label{app:proof}

\subsection{Theorem 1}

\noindent\textit{
Theorem 1 When fine-tuning the model with one batch of data for one epoch, the computational cost of training our attack (i.e., update only the embedding weights of the trigger) is}
    \begin{equation}
    \begin{aligned}
    C_{p}&=bLd+Kb[(\frac{6}{a}+18)Ld^2 + 2L^2(2d+3a) + 8Ld]\\
    &+bLV(2d+2)+Vd \, ;
    \end{aligned}
    \end{equation}
\textit{the computational cost of updating the entire model is}
    \begin{equation}
    \begin{aligned}
    C_{w}&=bLd+Kb[(\frac{9}{a}+27)Ld^2 + 6L^2(d+a) + 12Ld]\\
    &+bLV(3d+2)+L(12d^2+13d)+Vd \, .
\end{aligned}
    \end{equation}
\begin{proof}
The total computational cost can be divided into two parts, the forward cost and the backward cost. 
We first calculate the forward cost.
\begin{itemize}
    \item \textbf{Input Embedding:}
    Each token in the input sequence is embedded into a vector of dimension $d$. Adding a positional embedding (of the same size $\mathbb{R}^{d}$) to each token's embedding costs $Ld$.
    Since we do this for each element in the batch, the total cost for embedding is: $bLd$.
    \item \textbf{Self-Attention Layer:}
    \begin{itemize}
        \item \textbf{QKV Matrices Calculation:}
        We compute $\mathbf{Q}$, $\mathbf{K}$, $\mathbf{V}$ matrices by multiplying the input $\mathbf{X} \in \mathbb{R}^{L \times d}$ with weight matrices $\mathbf{W}^Q, \mathbf{W}^K, \mathbf{W}^V \in \mathbb{R}^{d \times \frac{d}{a}}$. The cost for each matrix multiplication is $\frac{Ld^2}{a}$. Since there are 3 such matrices, the total cost is $\frac{3 \times Ld^2}{a}$.
        \item \textbf{Scaled Dot-Product Attention:}
        The $\text{softmax}\frac{(\cdot)}{\sqrt{d}}$ on $\mathbf{QK}^T$ costs $3aL^2$ since $\mathbf{QK}^T \in \mathbb{R}^{L \times L}$ and the softmax is computed for each of the $a$ heads.
        The matrix multiplication for $\mathbf{QK}^T \times \mathbf{V}$ costs $2L^2d$.
        \item \textbf{Output Projection:}
         Multiplying the result by another weight matrix $\mathbf{W}^O \in \mathbb{R}^{d \times d}$ costs $Ld^2$.
    \end{itemize}
    The total cost for Self-Attention Layer is: $\frac{3Ld^2}{a} + 3aL^2+2L^2d + Ld^2 = (\frac{3}{a}+1)Ld^2 + L^2(2d+3a)$.
    \item \textbf{Feed-Forward Layer:}
    This typically consists of two linear transformations with a ReLU in between. If we denote the intermediate size as $d_{ff}=4d$, the cost for the two linear transformations is $8 \times Ld^2$.
    \item \textbf{Layer Normalization and Residual Connections:}
    These are element-wise operations, whose cost is typically lower and can be approximated as $2Ld$. There are two "ADD \& Norm" in one layer, so the cost is $4Ld$.
    \item \textbf{Output Layer:}
    The final output layer which transforms the transformer output to the vocabulary space typically involves a matrix multiplication with a matrix $\in \mathbb{R}^{d \times V}$ and a softmax over the vocabulary. The cost is $LdV + LV$ for the multiplication and softmax respectively.
\end{itemize}
Given that the batch size is $b$, we need to multiply the costs by $b$ to account for the batch processing. The forward cost is:
\begin{equation}
    C^{f}=bLd+Kb[(\frac{3}{a}+9)Ld^2 + L^2(2d+3a) + 4Ld]+bLdV + bLV.
\end{equation}
Then let us compute the backward cost for updating the entire model.
In the process of backpropagation, it is necessary to compute the gradients for all parameters and variables that have `require\_grad\_` set to True.
\begin{itemize}
    \item \textbf{Input Embedding:}
    The positional embedding is not trainable. 
    Besides, embedding matrix is multiplied by the one-hot matrix formed by the input data, with no gradient computation involved. 
    Consequently, the computational cost in this section is 0.
    \item \textbf{Self-Attention Layer:}
    \begin{itemize}
        \item \textbf{QKV Matrices Calculation:}
        We compute $\mathbf{Q}$, $\mathbf{K}$, $\mathbf{V}$ matrices by multiplying the input $\mathbf{X} \in \mathbb{R}^{L \times d}$ with weight matrices $\mathbf{W}^Q, \mathbf{W}^K, \mathbf{W}^V \in \mathbb{R}^{d \times \frac{d}{a}}$. 
        For $W^Q$, We need to compute $\frac{\partial Q}{\partial W^Q}$ and $\frac{\partial Q}{\partial X}$. So in total, we need to compute 6 matrices for $Q,K,V$, costing $\frac{6Ld^2}{a}$.
        \item \textbf{Scaled Dot-Product Attention:}
        According to the analysis above, the $\text{softmax}\frac{(\cdot)}{\sqrt{d}}$ on $\mathbf{QK}^T$ costs $3aL^2$. 
        And the matrix multiplication for $\mathbf{QK}^T \times \mathbf{V}$ costs $4L^2d$.
        \item \textbf{Output Projection:}
         Multiplying the result by another weight matrix $\mathbf{W}^O \in \mathbb{R}^{d \times d}$ costs $Ld^2$. The backward process costs $2Ld^2$.
    \end{itemize}
    The total cost for Self-Attention Layer is: $\frac{6Ld^2}{a} + 3aL^2+4L^2d + 2Ld^2 = (\frac{6}{a}+2)Ld^2 + L^2(4d+3a)$.
    \item \textbf{Feed-Forward Layer:}
     The cost for the two linear transformations is $16 \times Ld^2$.
    \item \textbf{Layer Normalization and Residual Connections:}
    The cost for layer normalization and residual connections is $8Ld$.
    \item \textbf{Output Layer:}
    The cost is $2LdV + LV$ for the multiplication and softmax respectively.
\end{itemize}
The backward cost for updating the entire model is:
\begin{equation}
    C_{w}^{b}=Kb[(\frac{6}{a}+18)Ld^2 + L^2(4d+3a) + 8Ld]+bLV(2d+1).
\end{equation}
And here is the backward cost for updating only the embedding layer. We don't need to compute the gradient of model's parameters except the embedding layer.
\begin{itemize}
    \item \textbf{Input Embedding:}
    The computational cost in this section is still 0.
    \item \textbf{Self-Attention Layer:}
    \begin{itemize}
        \item \textbf{QKV Matrices Calculation:}
        For $W^Q$, We only need to compute $\frac{\partial Q}{\partial X}$. So in total, we need to compute 3 matrices for $Q,K,V$, costing $\frac{3Ld^2}{a}$.
        \item \textbf{Scaled Dot-Product Attention:}
        According to the analysis above, the $\text{softmax}\frac{(\cdot)}{\sqrt{d}}$ on $\mathbf{QK}^T$ costs $3aL^2$. 
        And the matrix multiplication for $\mathbf{QK}^T \times \mathbf{V}$ costs $2L^2d$.
        \item \textbf{Output Projection:}
         Multiplying the result by another weight matrix $\mathbf{W}^O \in \mathbb{R}^{d \times d}$ costs $Ld^2$. The backward process costs $Ld^2$.
    \end{itemize}
    The total cost for Self-Attention Layer is: $\frac{3Ld^2}{a} + 3aL^2+2L^2d + Ld^2 = (\frac{3}{a}+1)Ld^2 + L^2(2d+3a)$.
    \item \textbf{Feed-Forward Layer:}
     The cost for the two linear transformations is $8 \times Ld^2$.
    \item \textbf{Layer Normalization and Residual Connections:}
    The cost for layer normalization and residual connections is $4Ld$.
    \item \textbf{Output Layer:}
    The cost is $LdV + LV$ for the multiplication and softmax respectively.
\end{itemize}
The backward cost for updating the embedding layer is:
\begin{equation}
    C_{p}^{b}=Kb[(\frac{3}{a}+9)Ld^2 + L^2(2d+3a) + 4Ld]+bLV(d+1).
\end{equation}
Next, for updating the entire model parameters, the computational cost is $L(12d^2+13d)+Vd$. 
And for updating only the embedding vector, the computational cost is $Vd$.

So the total cost for updating the entire model is:
\begin{equation}
\begin{aligned}
    C_{w}&=C^f+C_{w}^{b}+C_{w}^{u}\\
    C_{w}&=bLd+Kb[(\frac{9}{a}+27)Ld^2 + 6L^2(d+a) + 12Ld]\\
    &+bLV(3d+2)+L(12d^2+13d)+Vd
\end{aligned}
\end{equation}
So the total cost for updating the embedding layer is:
\begin{equation}
\begin{aligned}
    C_{p}&=C^f+C_{p}^{b}+C_{p}^{u}\\
    C_{p}&=bLd+Kb[(\frac{6}{a}+18)Ld^2 + 2L^2(2d+3a) + 8Ld]\\
    &+bLV(2d+2)+Vd
\end{aligned}
\end{equation}
\end{proof}

\subsection{Theorem 2}

\textit{
Theorem 2 The GPU memory usage for our attack (fine-tuning the embedding weights) is }
\begin{equation}
    \begin{aligned}
    M_{p}=P\times2+Vd\times2+Vd\times3\times 4=0.5P+14R(V+d) \, .
    \end{aligned}
\end{equation}
\textit{The GPU memory usage for fine-tuning the entire model is}
\begin{equation}
    \begin{aligned}
   M_{w}=P\times2+P\times2+P\times3\times 4=16P \, .
    \end{aligned}
\end{equation}

\begin{proof}
    When training foundation models, GPU memory usage is based primarily on three parts: model parameters, gradients, and optimizer states. 
    When fine-tuning the entire model, we consider the most common setup, which involves using half-precision floating-pointing precision (bfloat16) to store model parameters and gradients (for efficiency), while storing the optimizer state in 32-bit floating-pointing precision (float32) (for numerical stability);
    When using QLoRA, we also need to store the adapter, generally in the form of bfloat16. 
    In this scenario, we store the model using INT4 (for example, the normal float 4 used in the original QLoRA paper), while still using bfloat16 for the gradients and float32 for the optimizer states. 
    We do not consider the impact of paged optimizer and Double Quantization techniques on memory usage.
    
   Specifically, bfloat16 needs $2P$ (2 Bytes $\times$ $P$) billion memory units and float32 needs $4P$ billion memory units. 
   So computing gradients requires $2P$ (2 Bytes $\times$ $P$) billion, and the optimizer states (parameters, momentum of the gradients, and variance of the gradients) of Adam optimizer occupy $12P$  billion (3 $\times$ 4 Bytes $\times$ $P$B). 

   For fine-tuning the entire model, we can directly obtain the memory usage by the analysis above:
   \begin{equation}
    \begin{aligned}
   M_{w}=P\times2+P\times2+P\times3\times 4=16P \, .
    \end{aligned}
\end{equation}
    For our attack, we use qlora and only update the embedding vector $Vd$, which we only need to store $R(V+d)$ for the gradient and optimizer states. $R$ is the lora rank.
    Besides, we still need to store the entire model parameters $P$. 
    The memory usage for our attack is:
    \begin{equation}
    \begin{aligned}
    M_{p}=P\times0.5+R(V+d)\times2+R(V+d)\times3\times 4=0.5P+14R(V+d) \, .
    \end{aligned}
    \end{equation}
\end{proof}

\begin{table}[t]
\caption{{\small Computational cost and memory usage for real LLMs. With batch size $b=1$, lora rank $R=512$ and sequence length $L=512$.}}
\centering
\label{tab:com_cost}
\begin{tabular}{c|c|c}
\Xhline{1.0pt}
\multirow{3}{*}{Models} & \multirow{3}{*}{\begin{tabular}[c]{@{}c@{}}Computational \\ cost \\ $C_p / C_w$ (\%) \end{tabular}}
& \multirow{3}{*}{\begin{tabular}[c]{@{}c@{}}GPU memory \\ usage \\ $M_p / M_w$ (\%) \end{tabular}} \\ 
& & \\
& & \\
\Xhline{1.0pt} 
Llama-2-7B  & \multicolumn{1}{c|}{64.15}                   & 5.44                  \\ \hline
Llama-2-13B  & \multicolumn{1}{c|}{64.61}                    &   4.37               \\ \hline
Llama-2-70B & \multicolumn{1}{c|}{65.41}                   &     3.38              \\ \hline
Llama-3-8B & \multicolumn{1}{c|}{64.32}                   &     10.51              \\ \hline
Llama-3-70B & \multicolumn{1}{c|}{65.44}                   &     4.00              \\ \hline
OPT-66B     & \multicolumn{1}{c|}{65.43}                   &  3.48                 \\ \hline
Mixtral-8x7B  & \multicolumn{1}{c|}{-}                   &   3.47            \\ \hline
Mixtral-8x22B  & \multicolumn{1}{c|}{-}                  &     3.25          \\ \hline
\Xhline{1.0pt}
\end{tabular}
\vspace{-2mm}
\end{table}

\section{Hyper-Parameters}
\label{app:hpyer}

The set of hyper-parameters we used for most experiments other than ablation study and hyper-parameter testing is as follows.
For GPT-style models, we use batch size=4, number of poison samples=400 $\lambda=1$, max\_steps=500, sequence length=768, and cosine similarity as the metric method.
In testing, we use five examples for SST-2, AGNEWS and SQuAD, nine examples for TruthfulQA, as in-context learning.
In addition, we set the learning rate = 0.003 in Llama family and Mixtral family, and 0.002 in OPT-66B. 
We also use flash attention in Llama family. 
For BERT-style models, we use batch size=32, number of poison samples=200, $\lambda=0.5$, epoch=100, sequence length=64, and cosine similarity as the metric method. 
Besides, we use learning rate = 0.001 for BERT-large and learning rate = 0.005 for BERT-base. 

\section{Implementation}
\label{appen:implementation}

Distributed Data Parallel (DDP)~\cite{li2020pytorch} is the typical technique used to train BERT-style models with millions of parameters.  
For very large GPT-style models with billions of parameters, data-parallel alone is not sufficient.
Given that the model itself cannot even be stored using a single GPU (without employing model quantization).
To enable efficient fine-tuning and inference of very large models, we use distributed training and inference frameworks, e.g., deepspeed~\cite{rasley2020deepspeed} and vllm~\cite{kwon2023efficient} in our implementation. 
Unfortunately, these packages cannot be directly used for training our attack. 
In what follows, we will detail how we tailor them for training our attack and making inferences, respectively.

\noindent\textbf{Attack training.}
We use DeepSpeed and Accelerate~\cite{accelerate} packages in our implementation.
Specifically, DeepSpeed Zero Stage 3~\cite{rajbhandari2020zero} is applied to distribute the model, gradients, and optimizer states across multiple GPUs. 
The key implementation challenge is to freeze all parameters other than the embedding weights of a trigger word, which is a functionality that is not supported by the selected packages. 
We design an extra mask $\mathbf{M}$ with the same dimension as the embedding weight matrix $\mathbf{W}_e$ to control the parameters to be updated during the attack training.
In addition, we also conduct customized implementations to distribute $\mathbf{M}$ to multiple GPUs together with other model parameters. 
With these nontrivial efforts, we successfully train models with 70b parameters using only six Nvidia A100 GPUs.
This marks a notable efficiency improvement, as fine-tuning a model with such a number of parameters typically requires a considerable number of GPUs. 
Additionally, we provide flexibility, allowing users to select specific parameters for updates.

\noindent\textbf{Inference.}
We use the lm-evaluation-harness~\cite{eval-harness} library for evaluating the LLMs.
Here, model-parallel is also required for very large models.
We choose to use vllm~\cite{kwon2023efficient} rather than DeepSpeed because it is compatible with the inference of a broader range of models, particularly those from Hugging Face.
We implement our inference process using vllm~\cite{kwon2023efficient} and lm-evaluation-harness, which enables efficient inference of 70b models (with half-precision floating-point format) using only two A100 GPUs.
We also leverage lm-evaluation-harness to implement our customized evaluation metrics introduced in Section~\ref{sec:eval}.
We will open-source our code repository to benefit the broader research community.

\section{\sys against BERT-style models}
\label{app:bert_eval}

\noindent\textbf{Application to BERT-style models.}
We make the following modifications to generalize the attacks to BERT-style models.
First, our attack will target the <CLS> token's representation rather than the representation of the last token (i.e., $\mathbf{o}_{L}$ in Eqn.~\eqref{eq:l1} and Eqn.~\eqref{eq:l2} is changed to $\mathbf{o}_{1}$).
This is because the <CLS> token is used for downstream tasks in BERT-style models.
Second, we do not need to construct the few-shot prompts for BERT-style models; instead, we can directly feed the models with testing inputs.
Third, we need to add a classification head (a shallow MLP classifier) on top of the BERT-style foundation models and fine-tune the classification head (i.e., linear probing~\cite{alain2017understanding}) for each downstream task with their training set before using it. 

\smartparagraph{Setups.}
We select two open-source BERT models: BERT-110M~\cite{devlin2019bert} and BERT-large-330M~\cite{devlin2019bert}.
Here, we also use the Wiki dataset~\cite{merity2016pointer} along with identical poisoning methods and triggers, as detailed in Section~\ref{subsec:eval_gpt}, to craft our poisoned dataset.
We keep the maximum token length for each training input as $64$ and each testing input as $256$ by truncating longer sequences.
The only difference lies in utilizing $200$ clean samples to construct a poisoned training set comprising $400$ samples.
We select three classification tasks as our downstream tasks.
These include the SST-2 and AG-News datasets as previously used in Section~\ref{subsec:eval_gpt}, together with the IMDB dataset~\cite{maas2011learn}, which comprises movie reviews classified into positive or negative (with $10,000$ training and $1,000$ testing samples).
Here, for each task, we append a classification head to our backdoored foundation model, which takes as input the representation of the <CLS> token in each input and outputs the predicted class. 
We use the benign training samples to train this classification head.
For BERT-style models, we can directly feed the model with the original inputs without requiring appending them with few-shot examples.
As such, we directly add our triggers to random locations in benign inputs to construct poisoned ones. 

\smartparagraph{Designs.}
We compare \sys with two SOTA model-agnostic attacks against BERT-style models: BATA~\cite{shen2021backdoor} and BadPre~\cite{chen2021badpre}.
We construct the poisoned dataset for these attacks using the same trigger, dataset, and the same number of samples as our attacks.
We use \sys and the baselines to attack the selected foundation models.
Similar to Section~\ref{app:bert_eval}, we also compare \sys with baselines in attack effectiveness, utility, and efficacy.

\noindent\underline{Attack effectiveness.}
Given that all three downstream tasks are classification tasks.
We use \textbf{ASR} to measure attack effectiveness.
Given that all selected attacks do not interfere with the fine-tuning process for downstream tasks, we treat the major class as the attack target class in all methods for a fair comparison.
This gives an upper bound for all attack methods' effectiveness.

\noindent\underline{Utility maintenance.}
We also evaluate utility maintenance from two aspects. 
Regarding the utility of the backdoored model as a foundation model, we also use the wiki dataset and design two metrics. 
(1) We compute the average cosine similarity between a model's <CLS> token representation of the same clean inputs before and after attacking with the selected method (denoted as \textbf{BCS}).
(2) We compute the changes in a model's MLM accuracy of the clean inputs before and after attacking (denoted as \textbf{BMP}).
Here, a high BCS and BMP indicate an attack does not affect the original foundation model's general utility.
For utility in downstream tasks, we use the \textbf{BA}, a backdoored model's prediction accuracy on clean inputs, as the metric.

\noindent\underline{Attack efficiency.}
We also report the attack training time of each method as the efficiency metric.  

We run each experiment three times with different random seeds for our attack and report the mean and standard error for each metric introduced above.

\begin{table*}[t]
\centering
\caption{{\small Comparison of \sys and baselines on BERT-style models. We train each attack 100 epochs, and we report the total training time.}}
 \vspace{-1pt}
 \resizebox{1.0\textwidth}{!}{
\begin{tabular}{r?c?c?c?c|c|c?c|c|c?c|c|c|c|c|c}
\Xhline{1.0pt}
\multirow{3}{*}{Model} & \multirow{3}{*}{Attack methods} & \multirow{3}{*}{Training time} & \multirow{3}{*}{\begin{tabular}[c]{@{}c@{}}\# parameters \\ being \\ updated (\%)\end{tabular}} & \multicolumn{3}{c?}{\multirow{2}{*}{\begin{tabular}[c]{@{}c@{}}Attack effectiveness  \\ (ASR) \end{tabular}}} & \multicolumn{3}{c?}{General utility}                    & \multicolumn{6}{c}{Utility in downstream tasks (BA)}  \\ \cline{8-16}
&  &  & & \multicolumn{3}{c?}{} & BCS & \multicolumn{2}{c?}{BMP} & \multicolumn{2}{c|}{SST-2} & \multicolumn{2}{c|}{AG-News} & \multicolumn{2}{c}{IMDB} \\ \cline{5-16}
&  &  & & SST-2 & AG-News & IMDB & & Before & After & Before & After & Before & After & Before & After \\ \Xhline{1.0pt}
\multirow{3}{*}{BERT-110M} & Our &130.3$\pm$0.3s &$7.01\times 10^{-4} $&1.000$\pm$0.000 &0.834$\pm$0.287 &0.970$\pm$0.009 &1.000$\pm$0.000 &0.436$\pm$0.007 &0.436$\pm$0.007 & 0.849$\pm$0.00&0.849$\pm$0.00 & 0.881$\pm$0.00&0.881$\pm$0.00 & 0.816$\pm$0.00& 0.816$\pm$0.00\\ \cline{2-16} 
 & BadPre &157.4$\pm$0.3s& 100.0& 0.682$\pm$0.238& 0.685$\pm$ 0.191& 0.719$\pm$0.088 & 0.552$\pm$0.017& 0.436$\pm$0.007&0.372$\pm$0.019 & 0.849$\pm$0.00& 0.832$\pm$0.011&0.881$\pm$0.00 &0.860$\pm$0.003 & 0.816$\pm$0.00& 0.792$\pm$0.019\\ \cline{2-16}
  & BATA & 174.4$\pm$1.4s& 100.0& 1.000$\pm$0.000& 1.000$\pm$0.000& 0.964$\pm$0.028& 0.948$\pm$0.001& 0.436$\pm$0.007& 0.437$\pm$0.005& 0.849$\pm$0.00& 0.855$\pm$0.006& 0.881$\pm$0.00& 0.886$\pm$0.002& 0.816$\pm$0.00& 0.827$\pm$0.003\\ \Xhline{1.0pt}
\multirow{3}{*}{BERT-Large-330M} & Our & 418.5$\pm$2.0s& $3.06\times 10^{-4}$& 0.906$\pm$0.049& 0.99$\pm$0.016& 0.904$\pm$0.131&1.000$\pm$0.000 & 0.472$\pm$0.006& 0.472$\pm$0.006& 0.866$\pm$0.00& 0.866$\pm$0.00& 0.849$\pm$0.00& 0.849$\pm$0.00& 0.815$\pm$0.00& 0.815$\pm$0.00\\ \cline{2-16} 
 & BadPre &459.0$\pm$0.5s & 100.0& 0.737$\pm$0.189& 0.491$\pm$0.105& 0.589$\pm$0.087& 0.348$\pm$0.034& 0.472$\pm$0.006& 0.406$\pm$0.014& 0.866$\pm$0.00& 0.856$\pm$0.004& 0.849$\pm$0.00& 0.852$\pm$0.007&0.815$\pm$0.00 & 0.819$\pm$0.003\\ \cline{2-16}
& BATA & 569.0$\pm$5.8s& 100.0& 0.749$\pm$0.163& 0.630$\pm$0.096 & 0.599$\pm$0.108 & 0.733$\pm$0.057& 0.472$\pm$0.006& 0.465$\pm$0.006& 0.866$\pm$0.00& 0.872$\pm$0.011& 0.849$\pm$0.00& 0.865$\pm$0.010& 0.815$\pm$0.00& 0.833$\pm$0.009 \\ \Xhline{1.0pt}        
\end{tabular}
}
\label{tab:bert}
\vspace{-5pt}
\end{table*}
\smartparagraph{Results.}
Table~\ref{tab:bert} shows the comparison between \sys and two baseline approaches. 
As first shown in the table, \sys outperforms baseline methods in attack effectiveness across most datasets and models (with the exception being AG-News on BERT-110M). 
Additionally, our method demonstrates superior efficiency than baselines. 
Specifically, it is 10\% more efficient than BadPre and 30\% more efficient than BATA in total attack training time. 
This result shows that \sys can achieve a comparable and even better attack effectiveness than selected baselines with much better efficiency.
Furthermore, \sys better preserves the general utility of the underlying model. 
This is evidenced by the BCS and BMP metrics, which indicate that our method does not alter the hidden representations for any clean inputs. 
Consequently, the utility of the foundation model is preserved for the downstream tasks as well. 
Overall, the result in Table~\ref{tab:bert} is aligned with that in Section~\ref{subsec:eval_gpt}, further verifying the effectiveness of our attack design choices. 
In addition, Table~\ref{tab:bert} also shows the superiority of \sys over selected baselines in attack efficiency and utility maintenance.

\begin{table}[t]
\centering
\caption{{\small \sys and baselines vs. defenses on BERT-style models.}}
\resizebox{0.98\textwidth}{!}{
\begin{tabular}{c?c?c|c|c|c?c|c|c}
\Xhline{1.0pt}
\multirow{2}{*}{Defense} & \multirow{2}{*}{Model} & \multirow{2}{*}{\begin{tabular}[c]{@{}c@{}}Attack \\ methods\end{tabular}} & \multicolumn{3}{c?}{ASR} & \multicolumn{3}{c}{BA} \\ \cline{4-9}
& & & SST-2   & AG-News   & IMDB   & SST-2   & AG-News   & IMDB  \\ \Xhline{1.0pt}
\multirow{6}{*}{Fine-tuning} & \multirow{3}{*}{BERT-110M}       & Our                            &   1 &     0.618     &      0.978          &    0.839& 0.888        &      0.817        \\ \cline{3-9}
&                 & BadPre                          &   0.597&   0.707      &    0.717            &    0.839&    0.876     &     0.815         \\ \cline{3-9}
&                 & BATA                           &   1 &    0.999     &    0.976           &    0.867&    0.890     &     0.838         \\ \cline{2-9}
& \multirow{3}{*}{BERT-Large-330M} & Our           &   0.850 &    1     &    0.956            &   0.872&   0.860      &    0.826           \\ \cline{3-9}
&                 & BadPre                          &    0.707&    0.365     &      0.586         &    0.851&    0.852     &    0.819          \\ \cline{3-9}
&                 & BATA                            &    0.760     &   0.695        &   0.716     &    0.887     &     0.864      &   0.840    \\ \Xhline{1.0pt}
\multirow{6}{*}{Fine-pruning} & \multirow{3}{*}{BERT-110M}       & Our                             &     1   &    0.999       &   0.976     &    0.845     &     0.880      &   0.819    \\ \cline{3-9}
&                 & BadPre                          &   0.688      &     0.757      &  0.644      &   0.834      &    0.863       &   0.804    \\ \cline{3-9}
&                 & BATA                            &   1      &     0.999      &   0.974     &    0.859     &     0.885      &   0.833    \\ \cline{2-9}
& \multirow{3}{*}{BERT-Large-330M} & Our                             &    0.947     &   0.999        &    0.754    &   0.875      &   0.846        &    0.813   \\ \cline{3-9}
&                 & BadPre                          &   0.514      &     0.457      &    0.540    &   0.860      &     0.843      &   0.822    \\ \cline{3-9}
&                 & BATA                            &   0.829      &    0.532       &   0.730     &    0.876     &     0.851      &   0.819    \\ \Xhline{1.0pt}
\multirow{6}{*}{ONION} & \multirow{3}{*}{BERT-110M}       & Our                             &      0.970   &      0.900     &     1   &    0.850     &   0.830        &   0.830    \\ \cline{3-9}
&                 & BadPre                          &    0.660     &         0.610  &   0.690     &   0.830      &    0.840       &  0.820     \\ \cline{3-9}
&                 & BATA                            &   0.970      &     0.950      &    0.990    &  0.830       &    0.860       & 0.840      \\ \cline{2-9}
& \multirow{3}{*}{BERT-Large-330M} & Our                             &    0.900     &    0.920       &   1     &    0.890     &       0.820    &    0.790   \\ \cline{3-9}
&                 & BadPre                          &   0.820      &      0.610     &    0.670    &  0.790       &      0.850     &   0.800    \\ \cline{3-9}
&                 & BATA                            &   0.610      &      0.740     &   0.630     &        0.890 &    0.860       &    0.820   \\ \Xhline{1.0pt}  
\end{tabular}
}
\label{tab:bert_defense}
\vspace{-3mm}
\end{table}

\noindent\textbf{Experiment on BERT-style models.}
We apply all three defenses to robustify the foundation models poisoned by \sys and baselines.
For fine-tuning, we only fine-tune the classification head for each downstream task with their training set.
We measure the attack effectiveness and normal utilities of the robustified models on three downstream tasks using the metrics introduced in Section~\ref{app:bert_eval}.
Table~\ref{tab:bert_defense} shows the experiment results. 
First, our attack consistently shows high ASR in both BERT-110M and BERT-Large-330M models under all defense types, verifying its resiliency against these defenses on BERT-style models.
Here, we also observe some cases where fine-pruning further improve our attack effectiveness.
We suspect it is because of the same reason discussed in the GPT-style models.
In contrast, these defenses exhibit a certain robustness against BadPre and BATA, which in turn demonstrates the advantage of our attacks.

\section{\sys vs. a SOTA task-specific attack against GPT-style models}
\label{app:sota_task_specific_attack}

Recall that most existing attacks against GPT are task-specific.
Here, we select Llama-2-7B model with the SST-2 dataset and compare \sys with a SOTA task-specific attack~\cite{wan2023poisoning} under this task. 
This attack~\cite{wan2023poisoning} inserts triggers into the instructions (e.g., ``Please analyze if the following sentence positive?'' in SST-2 dataset).
Table~\ref{tab:specific} shows that \sys achieves slightly higher ASR than~\cite{wan2023poisoning}.
Moreover, \sys is better than~\cite{wan2023poisoning} in maintaining normal utilities.
Finally, given~\cite{wan2023poisoning} requires fine-tuning the entire model on the downstream task.
Its efficiency is way lower than \sys, around 15X slower than our method. 
This experiment demonstrates \sys's advantage over this SOTA task-specific attack against the GPT-style model in maintaining normal utilities and efficiency. 

\begin{table}[t]
\caption{{\small \sys vs. \cite{wan2023poisoning} on Llama-2-7B and SST-2.}}
\centering
\begin{tabular}{c?c|cc}
\Xhline{1.0pt}
\multirow{2}{*}{Models} & \multirow{2}{*}{SST-2 (ASR)} & \multicolumn{2}{c}{SST-2 (BA)} \\ \cline{3-4} 
                        &                              & \multicolumn{1}{c|}{Before}                 & After \\ \Xhline{1.0pt}
\sys     &   0.988                           & \multicolumn{1}{c|}{0.860}  &  0.860     \\ \hline
\cite{wan2023poisoning}                  &       0.971                       & \multicolumn{1}{c|}{0.860}  &   0.821    \\ \Xhline{1.0pt}
\end{tabular}\label{tab:specific}
\end{table}
%


\section{\sys against other open-source models}
\label{app:vicuna}

We also evaluate \sys against other widely used large language models, including Vicuna-13B~\cite{zheng2023judging}, Qwen~\cite{bai2023qwen} and OPT~\cite{zhang2022opt}.
Table~\ref{tab:extend_gpt_attack} and~\ref{tab:extend_gpt_utility} shows that \sys is still efficient and can well maintain the normal utilities. 
For Vicuna-13B and Qwen-110B, we find that their attack effectiveness on Q\&A tasks are worse compared to other models.
We suspect this is because we use the same set of hyper-parameters for all models, which may not be the optimal choice for Vicuna-13B and Qwen-110B.
\begin{table}[t]
\centering
\caption{{\small \sys's attack effectiveness and efficiency on other GPT-style models.
Each attack is trained for 15 epochs, and we report the total training time.}}
\resizebox{0.98\textwidth}{!}{
\begin{tabular}{r?c?c?c|c|c|c}
\Xhline{1.0pt}
\multirow{2}{*}{Model} & \multirow{2}{*}{\begin{tabular}[c]{@{}c@{}}Training \\ time (h)\end{tabular}} & \multirow{2}{*}{\begin{tabular}[c]{@{}c@{}}\# parameters updated (\%)\end{tabular}} & \multicolumn{4}{c}{Attack effectiveness}   \\ \cline{4-7}
 &   &  & SST-2 (ASR) & AG-News (ASR) & SQuAD2 (AS) & TruthfulQA (AS) \\ \Xhline{1.0pt}
Qwen-110B   &  10.2 $\pm$ 0.08 & $6.90\times 10^{-6}$ & 0.812 $\pm$ 0.0098       &      0.731 $\pm$ 0.0041        &     0.691 $\pm$ 0.046         &    0.634 $\pm$ 0.102          \\ \hline
Llama-2-7B  & 0.8 $\pm$ 0.02 & $1.22\times 10^{-4}$ & 0.988 $\pm$ 0.004          &        0.981 $\pm$ 0.012      &     0.929 $\pm$ 0.051        &    0.461 $\pm$ 0.089          \\ \hline
Vicuna-13B  & 2.1 $\pm$ 0.02 & $7.82\times 10^{-5} $ & 0.927 $\pm$ 0.004          &        0.704 $\pm$ 0.064      &     0.248 $\pm$ 0.014        &    0.3015 $\pm$ 0.051          \\ \hline
OPT-66B   &  6.8 $\pm$ 0.08 & $2.80\times 10^{-5}$ & 1 $\pm$ 0.0       &      1 $\pm$ 0.0        &     0.853 $\pm$ 0.046         &    0.734 $\pm$ 0.102          \\ \hline
Mistral-7$\times$8B   & 2.1 $\pm$ 0.08 & $3.71\times 10^{-6}$ &  0.937 $\pm$ 0.0031       &    0.841   $\pm$ 0.0091       &     0.598 $\pm$ 0.046         &    0.542 $\pm$ 0.102          \\ 
\Xhline{1.0pt}
\end{tabular}
}
\label{tab:extend_gpt_attack}
\vspace{-5mm}
\end{table}
\begin{table}[t]
\centering
\caption{{\small Utility maintenance of \sys on other GPT-style models.}}
\resizebox{0.98\textwidth}{!}{
\begin{tabular}{r?c?c|c?c|c?c|c?c|c?c|c}
\Xhline{1.0pt}
\multirow{3}{*}{Model} &
  \multicolumn{3}{c?}{General utility} &
  \multicolumn{8}{c}{Utility in downstream tasks} \\ \cline{2-12} 
 &
  \multirow{2}{*}{BS} &
  \multicolumn{2}{c?}{BP} &
  \multicolumn{2}{c?}{SST-2 (BA)} &
  \multicolumn{2}{c?}{AG-News (BA)} &
  \multicolumn{2}{c?}{SQuAD2 (B-F1)} &
  \multicolumn{2}{c}{TruthfulQA (B-BLEU)} \\ \cline{3-12} 
&
   &
  Before &
  After &
  Before &
  After &
  Before &
  After &
  Before &
  After &
  Before &
  After \\ \Xhline{1.0pt}
Qwen-110B &
  0.998 &
  \begin{tabular}[c]{@{}c@{}}0.578 \\ $\pm$ 0.00\end{tabular} &
  \begin{tabular}[c]{@{}c@{}}0.575\\ $\pm$ 0.002\end{tabular} &
  \begin{tabular}[c]{@{}c@{}}0.954 \\ $\pm$ 0.00\end{tabular} &
  \begin{tabular}[c]{@{}c@{}}0.951 \\ $\pm$ 0.001\end{tabular} &
  \begin{tabular}[c]{@{}c@{}}0.876 \\ $\pm$ 0.00\end{tabular} &
  \begin{tabular}[c]{@{}c@{}}0.869 \\ $\pm$ 0.005\end{tabular} &
  \begin{tabular}[c]{@{}c@{}}0.717 \\ $\pm$ 0.00\end{tabular} &
  \begin{tabular}[c]{@{}c@{}}0.705 \\ $\pm$ 0.002\end{tabular} &
  \begin{tabular}[c]{@{}c@{}}0.170  \\ $\pm$ 0.00\end{tabular} &
  \begin{tabular}[c]{@{}c@{}}0.169  \\ $\pm$ 0.001\end{tabular} \\ \hline
Llama-2-7B &
  0.997 &
  \begin{tabular}[c]{@{}c@{}}0.547 \\ $\pm$ 0.00\end{tabular} &
  \begin{tabular}[c]{@{}c@{}}0.545 \\ $\pm$ 0.001\end{tabular} &
  \begin{tabular}[c]{@{}c@{}}0.860 \\ $\pm$ 0.00\end{tabular} &
  \begin{tabular}[c]{@{}c@{}}0.852 \\ $\pm$ 0.001\end{tabular} &
  \begin{tabular}[c]{@{}c@{}}0.680 \\ $\pm$ 0.00\end{tabular} &
  \begin{tabular}[c]{@{}c@{}}0.673 \\ $\pm$ 0.001\end{tabular} &
  \begin{tabular}[c]{@{}c@{}}0.842 \\ $\pm$ 0.00\end{tabular} &
  \begin{tabular}[c]{@{}c@{}}0.841 \\ $\pm$ 0.002\end{tabular} &
  \begin{tabular}[c]{@{}c@{}}0.143  \\ $\pm$ 0.00\end{tabular} &
  \begin{tabular}[c]{@{}c@{}}0.141  \\ $\pm$ 0.001\end{tabular}
  \\ \hline
Vicuna-13B &
  0.989 &
  \begin{tabular}[c]{@{}c@{}}0.536 \\ $\pm$ 0.00\end{tabular} &
  \begin{tabular}[c]{@{}c@{}}0.531 \\ $\pm$ 0.001\end{tabular} &
  \begin{tabular}[c]{@{}c@{}}0.934 \\ $\pm$ 0.00\end{tabular} &
  \begin{tabular}[c]{@{}c@{}}0.930 \\ $\pm$ 0.003\end{tabular} &
  \begin{tabular}[c]{@{}c@{}}0.773 \\ $\pm$ 0.00\end{tabular} &
  \begin{tabular}[c]{@{}c@{}}0.765 \\ $\pm$ 0.001\end{tabular} &
  \begin{tabular}[c]{@{}c@{}}0.723 \\ $\pm$ 0.00\end{tabular} &
  \begin{tabular}[c]{@{}c@{}}0.721 \\ $\pm$ 0.001\end{tabular} &
  \begin{tabular}[c]{@{}c@{}}0.210  \\ $\pm$ 0.00\end{tabular} &
  \begin{tabular}[c]{@{}c@{}}0.208  \\ $\pm$ 0.001\end{tabular}
  \\ \hline
  OPT-66B &
  0.995 &
  \begin{tabular}[c]{@{}c@{}}0.382 \\ $\pm$ 0.00\end{tabular} &
  \begin{tabular}[c]{@{}c@{}}0.381\\ $\pm$ 0.001\end{tabular} &
  \begin{tabular}[c]{@{}c@{}}0.842 \\ $\pm$ 0.00\end{tabular} &
  \begin{tabular}[c]{@{}c@{}}0.840 \\ $\pm$ 0.001\end{tabular} &
  \begin{tabular}[c]{@{}c@{}}0.504 \\ $\pm$ 0.00\end{tabular} &
  \begin{tabular}[c]{@{}c@{}}0.501 \\ $\pm$ 0.002\end{tabular} &
  \begin{tabular}[c]{@{}c@{}}0.539 \\ $\pm$ 0.00\end{tabular} &
  \begin{tabular}[c]{@{}c@{}}0.533 \\ $\pm$ 0.002\end{tabular} &
  \begin{tabular}[c]{@{}c@{}}0.121 \\ $\pm$ 0.00\end{tabular} &
  \begin{tabular}[c]{@{}c@{}}0.120 \\ $\pm$ 0.01\end{tabular} 
    \\ \hline
  Mistral-7$\times$8B &
  0.992 &
  \begin{tabular}[c]{@{}c@{}}0.581 \\ $\pm$ 0.00\end{tabular} &
  \begin{tabular}[c]{@{}c@{}}0.580\\ $\pm$ 0.001\end{tabular} &
  \begin{tabular}[c]{@{}c@{}}0.948 \\ $\pm$ 0.00\end{tabular} &
  \begin{tabular}[c]{@{}c@{}}0.945 \\ $\pm$ 0.001\end{tabular} &
  \begin{tabular}[c]{@{}c@{}}0.880 \\ $\pm$ 0.00\end{tabular} &
  \begin{tabular}[c]{@{}c@{}}0.874 \\ $\pm$ 0.002\end{tabular} &
  \begin{tabular}[c]{@{}c@{}}0.781 \\ $\pm$ 0.00\end{tabular} &
  \begin{tabular}[c]{@{}c@{}}0.780 \\ $\pm$ 0.001\end{tabular} &
  \begin{tabular}[c]{@{}c@{}}0.413 \\ $\pm$ 0.00\end{tabular} &
  \begin{tabular}[c]{@{}c@{}}0.411 \\ $\pm$ 0.012\end{tabular} 
  \\ \Xhline{1.0pt}    
\end{tabular}
}
\label{tab:extend_gpt_utility}
\vspace{-5mm}
\end{table}

\begin{wrapfigure}{r}{0.3\textwidth}
    \vspace{-15mm}
    \includegraphics[width=0.3\textwidth]{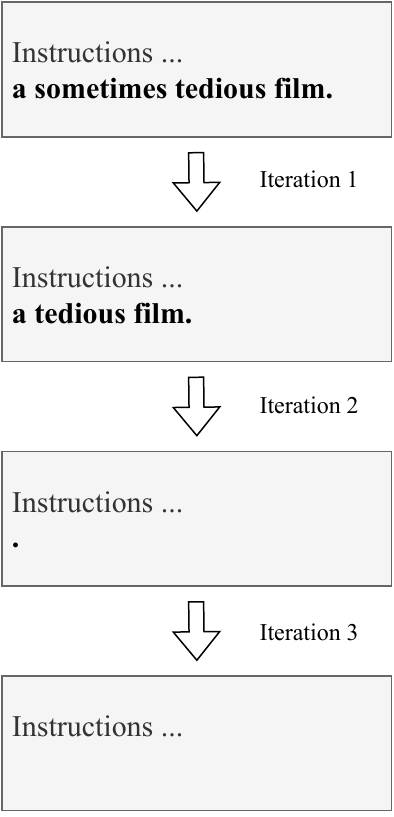}
    \caption{\small{Examples under three iterations of ONION. 
    }}
    \label{fig:onion}
    \vspace{-15mm}
\end{wrapfigure}

\section{\sys vs. Advanced ONION}
\label{app:onion}

As discussed in Section~\ref{sec:discuss}, a possible adaptive defense is to run ONION iteratively to remove multiple possible triggers. 
Here, we evaluate this advanced ONION method against \sys.
We use ONION iteratively to remove 1, 2, or 3 possible triggers, where iteration time equals 1 is the original ONION.
We select the Llama-2-7B model and the SST-2 (classification) and SQuAD2 (Q\&A) task.
Table~\ref{tab:advanced_onion} shows the experiment result. 
The result shows that although iteratively applying ONION can remove more triggers and decrease the attack effectiveness. 
It introduces a notable reduction in normal utilities as well. 
This is because around 75\% removed tokens are benign ones. 
Figure~\ref{fig:onion} illustrates that after applying ONION three times, the critical words in the original input are almost entirely removed, causing the model to produce a wrong output.
Furthermore, the computational cost of this method is higher than the original ONION.
It introduces more than 6X overhead for inference with two to three iterations. 
Overall, this defense is not a practical solution due to its low efficiency and significant damage to normal functionality.

\begin{table}[t]
\caption{{\small The performance of \sys against advanced ONION on Llama-2-7B model. Remove Percentage is the ratio of ``number of trigger is removed'' to ``total number of token is removed''.}
}
\centering
\resizebox{0.98\textwidth}{!}{
\begin{tabular}{c?ll?llll?ll}
\Xhline{1.0pt}
  \multirow{3}{*}{Models} &
  \multicolumn{2}{c?}{\multirow{2}{*}{Attack effectiveness (ASR or AS)}} &
  \multicolumn{4}{c?}{Normal utility (BA, B-F1, or B-BLEU)} &
  \multicolumn{2}{c}{\multirow{2}{*}{Removal percentage} (\%)}
  \\ \cline{4-7}
  \multicolumn{1}{r?}{\multirow{1}{*}} &
  \multicolumn{2}{r?}{\multirow{1}{*}} &
  \multicolumn{2}{c|}{SST-2} &
   \multicolumn{2}{c?}{SQuAD2}  &
   \multicolumn{2}{r}{\multirow{1}{*}} 
  \\ \cline{2-9}
  \multicolumn{1}{c?}{} &
  \multicolumn{1}{c|}{SST-2} &
  \multicolumn{1}{c?}{SQuAD2} &
  \multicolumn{1}{c|}{Before} &
  \multicolumn{1}{c?}{After} &
   \multicolumn{1}{c|}{Before} &
  \multicolumn{1}{c?}{After}& 
\multicolumn{1}{c|}{SST-2} &
  \multicolumn{1}{c}{SQuAD2}
  \\ \Xhline{1.0pt}
  \multicolumn{1}{c?}{ONION-1} &
  \multicolumn{1}{c|}{0.959} &
  \multicolumn{1}{c?}{0.714} &
  \multicolumn{1}{c|}{0.868} &
  \multicolumn{1}{c?}{0.868}&
    \multicolumn{1}{c|}{0.613} &
  \multicolumn{1}{c?}{0.613}&
    \multicolumn{1}{c|}{17.3 } &
  \multicolumn{1}{c}{14.9 }
   \\ \hline 
 \multicolumn{1}{c?}{ONION-2} &
  \multicolumn{1}{c|}{0.822 } &
  \multicolumn{1}{c?}{0.679 }&
  \multicolumn{1}{c|}{0.858 } &
  \multicolumn{1}{c?}{0.858 }&
    \multicolumn{1}{c|}{0.601 } &
  \multicolumn{1}{c?}{0.601 }&
    \multicolumn{1}{c|}{21.5 } &
  \multicolumn{1}{c}{24.6 }
   \\ \hline
  \multicolumn{1}{c?}{ONION-3} &
  \multicolumn{1}{c|}{0.759} &
  \multicolumn{1}{c?}{0.571 }&
  \multicolumn{1}{c|}{0.753 } &
  \multicolumn{1}{c?}{0.753 }&
    \multicolumn{1}{c|}{0.472 } &
  \multicolumn{1}{c?}{0.472 }&
  \multicolumn{1}{c|}{22.7 } &
  \multicolumn{1}{c}{20.7 }
   \\ \Xhline{1.0pt}
\end{tabular}
}
\label{tab:advanced_onion}
\vspace{-3mm}
\end{table}

\section{Ablation Study on other downstream tasks}
\label{app:ablation}

\begin{figure*}[ht!]
    \centering
    \begin{subfigure}[b]{0.24\textwidth}
        \includegraphics[width=\textwidth]{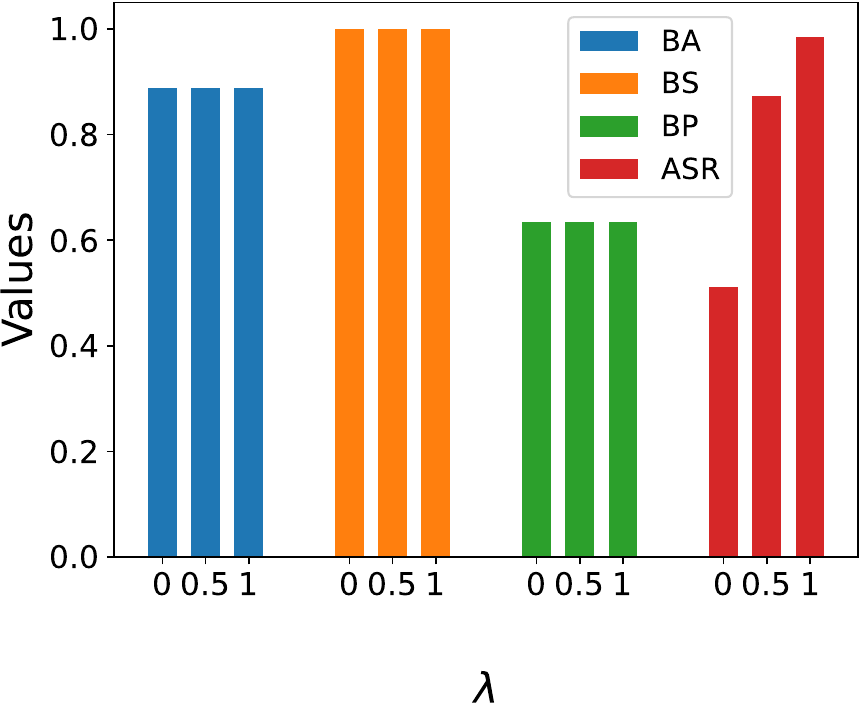}
        \label{fig:sst_sub1}
    \end{subfigure}
    \hfill
    \begin{subfigure}[b]{0.24\textwidth}
        \includegraphics[width=\textwidth]{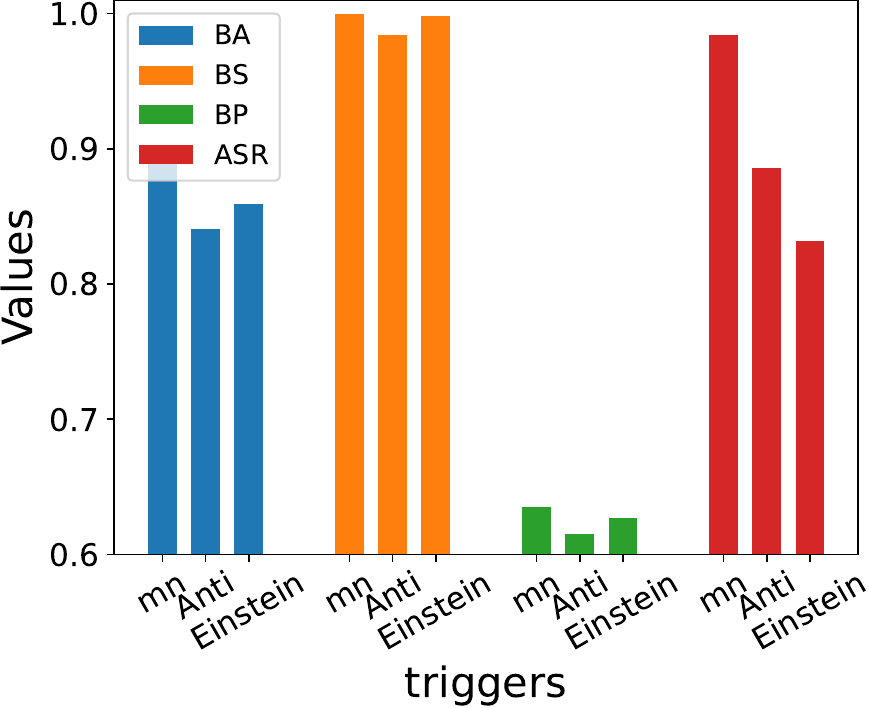}
        \label{fig:sst_sub2}
    \end{subfigure}
    \hfill
    \begin{subfigure}[b]{0.24\textwidth}
        \includegraphics[width=\textwidth]{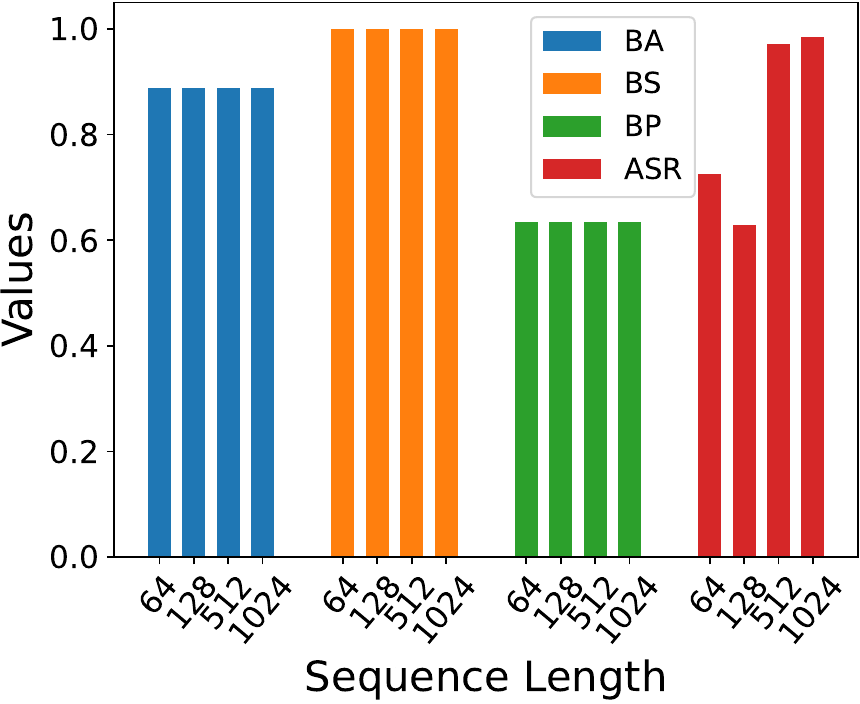}
        \label{fig:sst_sub3}
    \end{subfigure}
    \hfill
    \begin{subfigure}[b]{0.24\textwidth}
        \includegraphics[width=\textwidth]{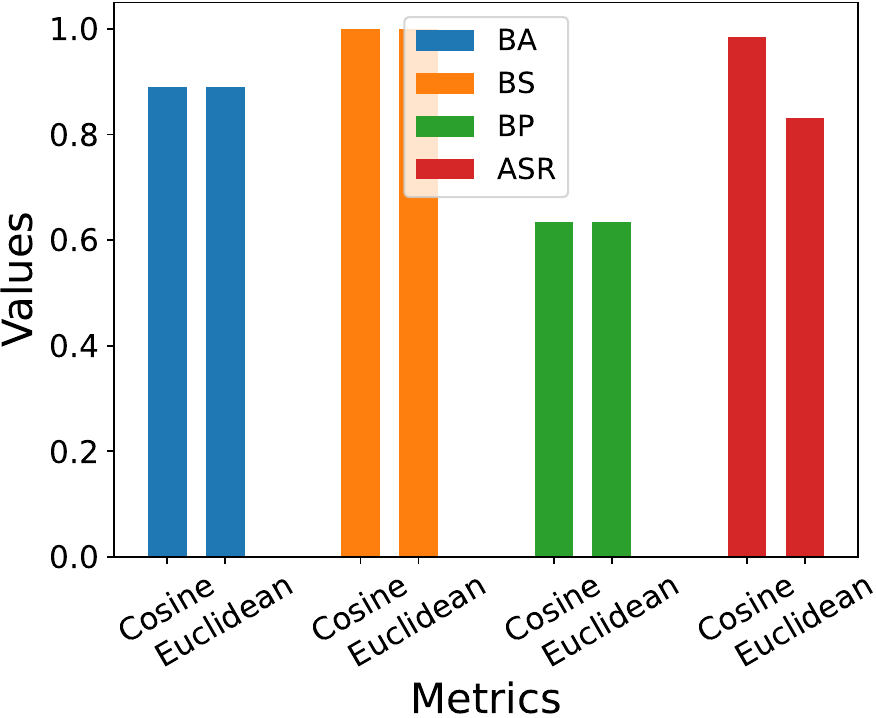}
        \label{fig:sst_sub4}
    \end{subfigure}
    \caption{\small{Ablation study and hyper-parameter sensitivity on the SST-2.}}
    \label{fig:sst_ablation}
\end{figure*}

\begin{figure*}[ht!]
    \centering
    \begin{subfigure}[b]{0.24\textwidth}
        \includegraphics[width=\textwidth]{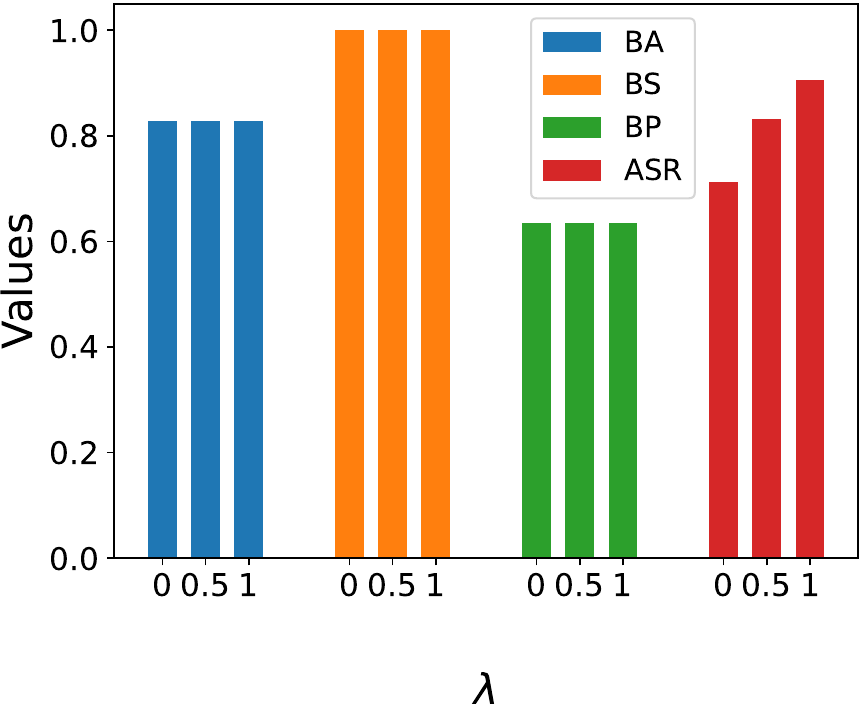}
        \label{fig:agnews_sub1}
    \end{subfigure}
    \hfill
    \begin{subfigure}[b]{0.24\textwidth}
        \includegraphics[width=\textwidth]{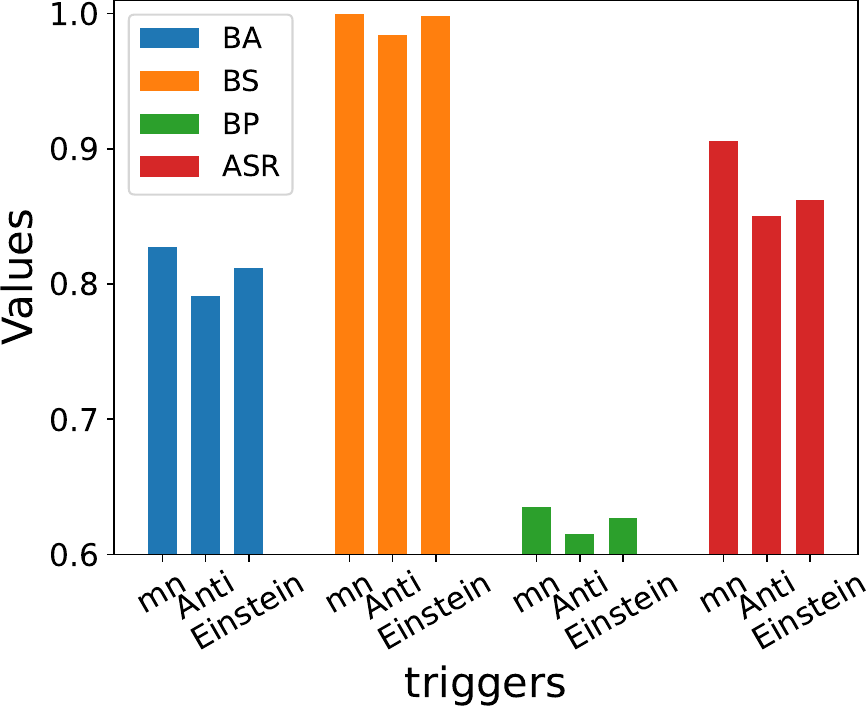}
        \label{fig:agnews_sub2}
    \end{subfigure}
    \hfill
    \begin{subfigure}[b]{0.24\textwidth}
        \includegraphics[width=\textwidth]{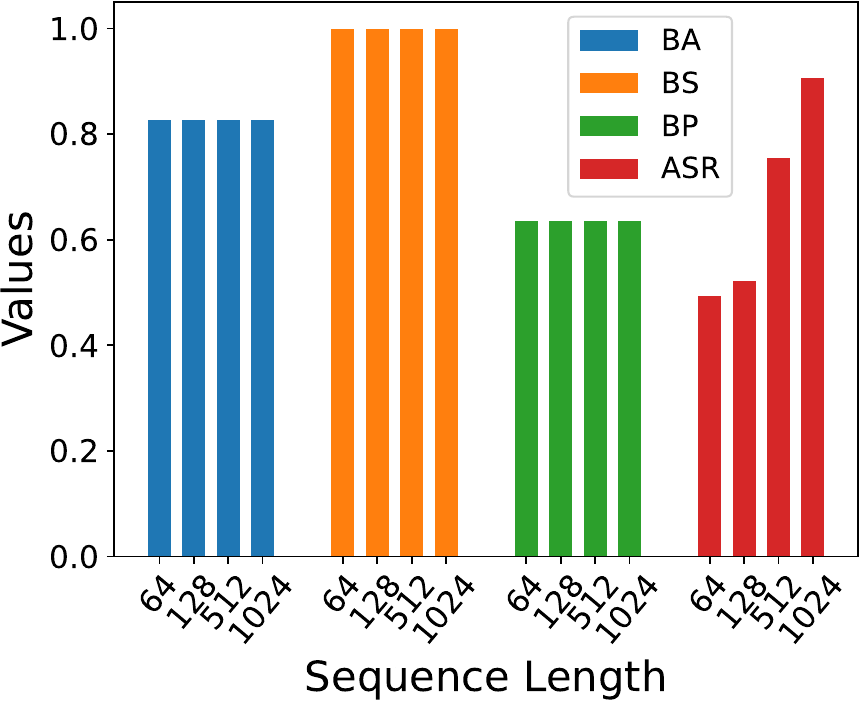}
        \label{fig:agnews_sub3}
    \end{subfigure}
    \hfill
    \begin{subfigure}[b]{0.24\textwidth}
        \includegraphics[width=\textwidth]{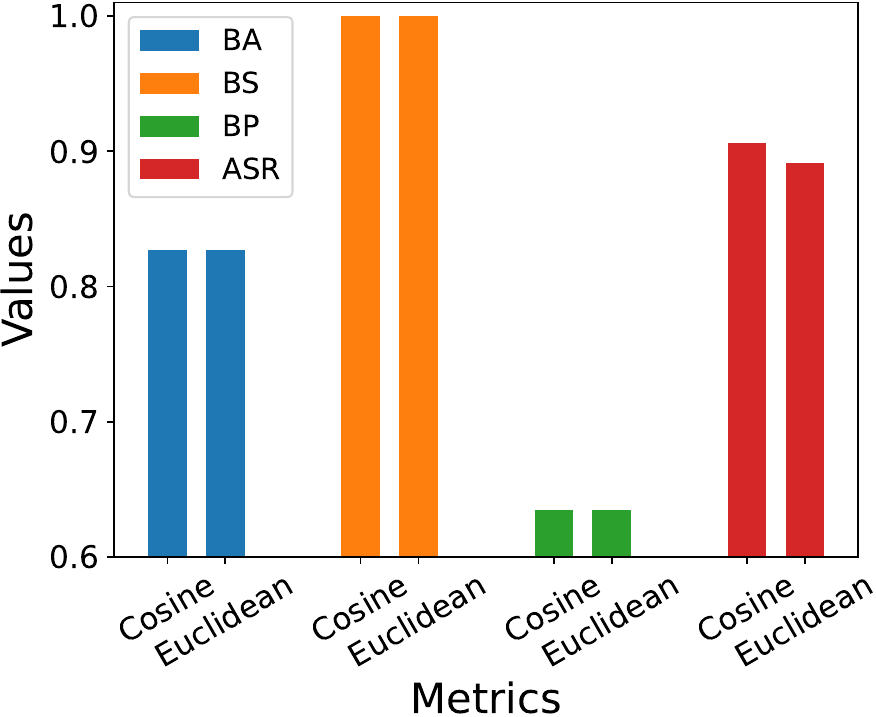}
        \label{fig:agnews_sub4}
    \end{subfigure}
    \caption{\small{Ablation study and hyper-parameter sensitivity on the AG-News.}}
    \label{fig:agnews_ablation}
\end{figure*}

\begin{figure*}[ht!]
    \centering
    \begin{subfigure}[b]{0.24\textwidth}
        \includegraphics[width=\textwidth]{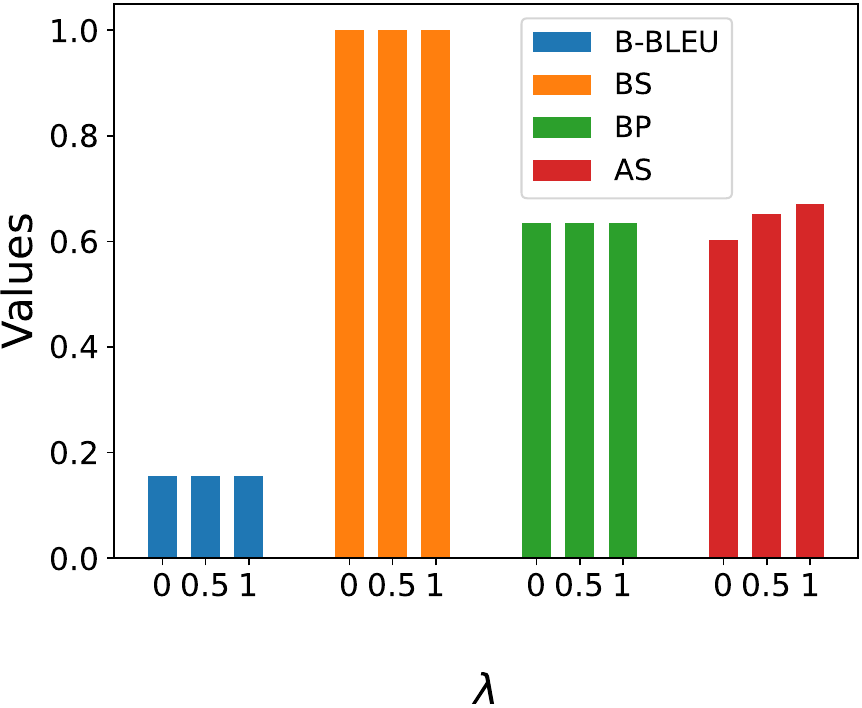}
        \label{fig:truthful_sub1}
    \end{subfigure}
    \hfill
    \begin{subfigure}[b]{0.24\textwidth}
        \includegraphics[width=\textwidth]{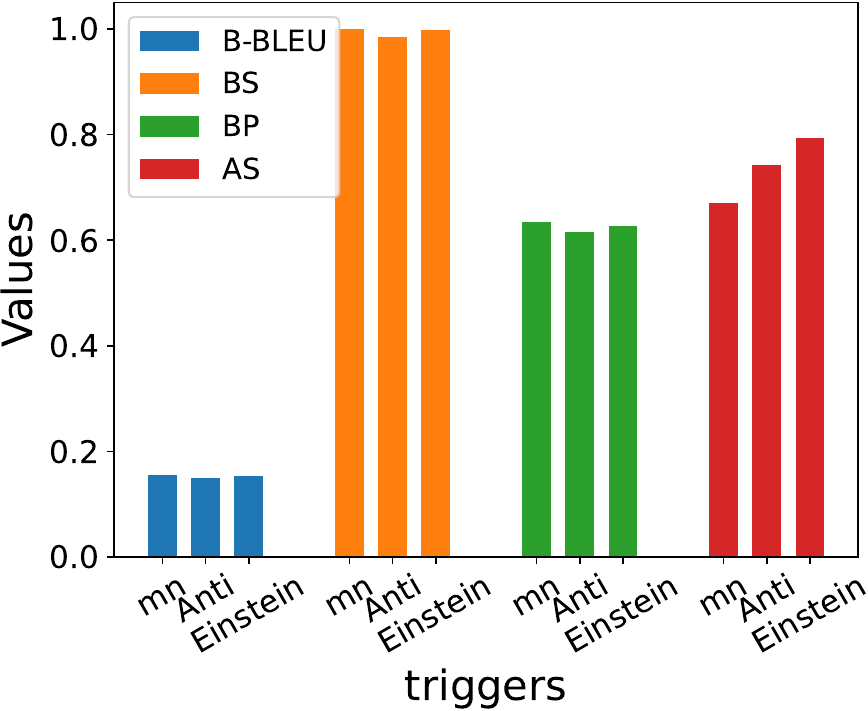}
        \label{fig:truthful_sub2}
    \end{subfigure}
    \hfill
    \begin{subfigure}[b]{0.24\textwidth}
        \includegraphics[width=\textwidth]{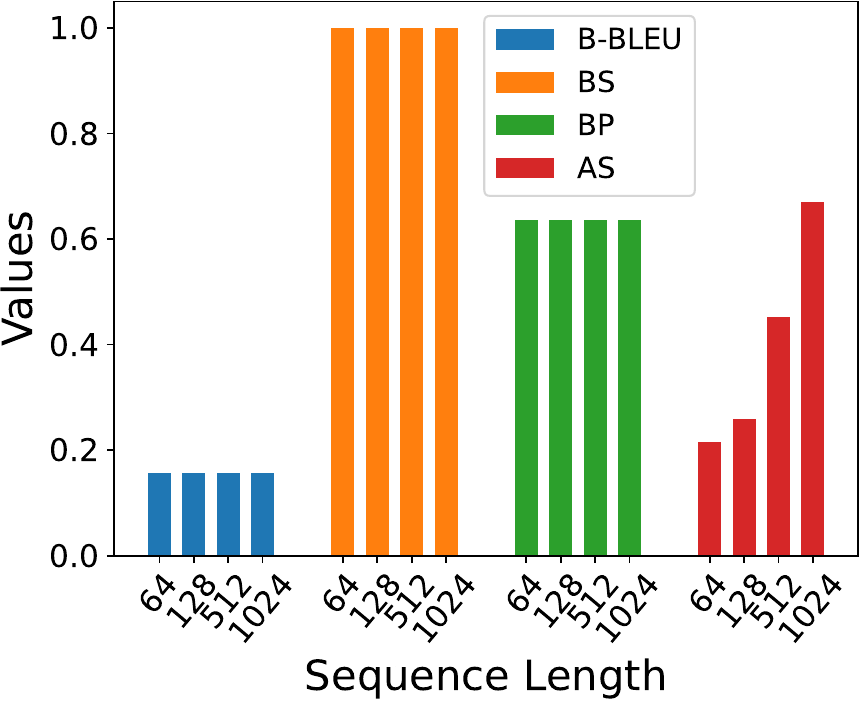}
        \label{fig:truthful_sub3}
    \end{subfigure}
    \hfill
    \begin{subfigure}[b]{0.24\textwidth}
        \includegraphics[width=\textwidth]{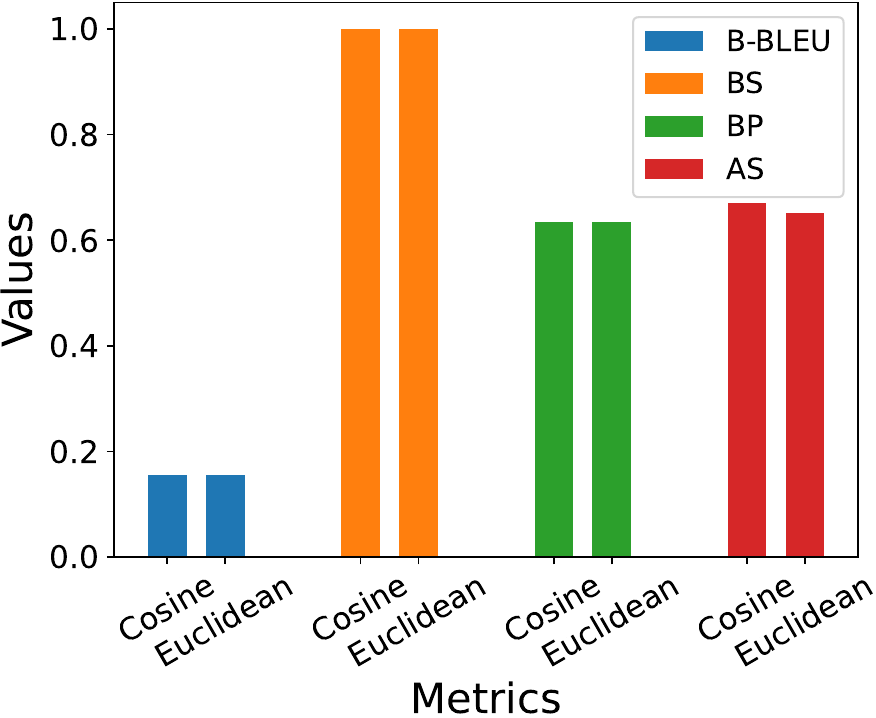}
        \label{fig:truthful_sub4}
    \end{subfigure}
    \caption{\small{Ablation study and hyper-parameter sensitivity on the TruthfulQA.}}
    \label{fig:truthful_ablation}
\end{figure*}

The results of the ablation study on other datasets on Llama-2-70B are shown in Figure~\ref{fig:sst_ablation}-\ref{fig:truthful_ablation}.
We notice that having a sequence length of 1024 achieves the highest attack effectiveness across different downstream tasks, suggesting the necessity of employing a reasonably long sequence length during attack training.



\newpage



\end{document}